\documentclass[aps,pre,showpacs,floats,12pt,nofootinbib
]{revtex4}
\usepackage{graphicx}
\usepackage{amsfonts}
\usepackage{txfonts} 
\usepackage{times}
\usepackage{subfigure}
\newcommand{\beq}{\begin{equation}}     \newcommand{\eeq}{\end{equation}}
\newcommand{\beqa}{\begin{eqnarray}}    \newcommand{\eeqa}{\end{eqnarray}}
\newcommand{\bde}{\begin{description}}  \newcommand{\ede}{\end{description}}
\newcommand{\ben}{\begin{enumerate}}    \newcommand{\een}{\end{enumerate}}

\newcommand{\kT}{{k_{\rm B}T} } 

\newcommand{\bm}[1]{\mbox{\boldmath ${#1}$}}
\newcommand{\citer}[1]{{Eq.\ref{#1}}}

\newcommand{\rmd}{{\rm d}}

\newcommand{\inRbracket}[1]{{\left({#1}\right)}}

\begin{document}
\title{Compatibility between itinerant synaptic receptors\\
and stable postsynaptic structure
} 
\author{ Ken Sekimoto$^{*1,*2}$ and  
Antoine Triller$^{*3}$
} 
\affiliation{
${}^{*1}$ {
Mati\`{e}res et Syst\`{e}mes Complexes, CNRS-UMR7057, Universit\'e
  Paris 7, France
} \\
${}^{*2}$ { {Gulliver,} CNRS-UMR7083, ESPCI, 
Paris, France
} \\
${}^{*3}$ {INSERM, U789, Biologie Cellulaire de la Synapse N\&P, Ecole Normale Sup\'erieure, Paris, France }}

\begin{abstract}

\noindent
The density of synaptic receptors in front of presynaptic release sites 
is stabilized in the presence of scaffold proteins,
but the receptors and scaffold molecules have local exchanges with characteristic times shorter than that of the receptor-scaffold assembly. 
We propose a mesoscopic model to account for the regulation of the local density of receptors as quasi-equilibrium.
It is based on two zones (synaptic and extrasynaptic) and multi-layer (membrane, sub-membrane and cytoplasmic) topological organization.
 The model includes the balance of chemical potentials associated with the receptor and scaffold protein concentrations in the various compartments.
The model shows highly cooperative behavior including a ``phase change'' resulting in the formation of well-defined post-synaptic domains. 
 This study provides {theoretical tools} 
  to approach the complex issue of synaptic stability at the synapse, where receptors are transiently trapped yet rapidly diffuse laterally on the plasma membrane. 
\end{abstract}

\maketitle

\section{Introduction - Biological background and problem} 
\label{sec:introduction}
A large body of structural data has shown that {synaptic} receptors
accumulate in the postsynaptic density 
(PSD).  
The classic static view of receptor distribution 
was challenged a few years ago by the evidence that receptor numbers at synapses are tuned during 
regulation of synaptic strength  (reviewed in Refs \cite{Malinov02,ShengKim,Bredt03} ). 
This is now considered one of the molecular bases of synaptic plasticity.
{Synaptic plasticity is one of the most commonly used concepts to explain the capacity of the 
brain to adapt to external and internal conditions and to modify the properties of neuronal networks in relation to development and learning. 
The tuning of receptor numbers} has led to the important notion of receptor flux into and out  
of synapses, both at rest and during plasticity. 
It has prompted the development of dynamic real-time imaging approaches in
living neurons, such as video-microscopy of green fluorescent protein  
(GFP)-tagged receptors, to go beyond the fixed snapshots given by immuno-cytochemistry. 
However, these multimolecular approaches have limits: 
Although they can detect receptor fluxes (e.g. 
using fluorescence recovery after photobleaching, FRAP), in basal conditions when synaptic receptor numbers remain constant overall, they 
cannot monitor minute exchanges between compartments. 
The advent of single molecule imaging techniques now enables measurement of individual receptor movements in identified sub-membrane compartments, and reveals the inhomogeneities and new physical parameters important for the understanding of {\it receptor trafficking}. 
The chemical approach is appropriate to further clarify 
the interplay between 
 the constituent molecules of the postsynaptic molecular assembly.
Our theoretical 
  model is intended to present a realistic view of how those molecules
  behave both individually and collectively.  

The synapse as a multimolecular assembly should be viewed as a 
construction 
where the constituent elements are characterized by dwell 
 time (local turnover). 
In other words, the synapse as a whole and the constituent elements have specific characteristic 
times. 
This view is not unique to the synapse, but is now well accepted for structures like actin and 
microtubules with well-known tread-milling behavior or the turnover of ATPase molecular motors 
during cell motility\cite{verkhovsky} and in intracellular 
trafficking \cite{rev-kinesin}.
 Theoretical frameworks accounting for 
the dynamics of these structures have been proposed and have allowed the development of a new
experimental paradigm \cite{brisure,kerato5,kerato5bis}.
Such a theoretical approach has been lacking for the postsynaptic 
membrane. 
 The structures of the synapse that are unified for excitatory and inhibitory
 contacts have been extensively studied during the two last decades.
The recent development of dynamic methods and real-time imaging, e.g. 
single-particle tracking (SPT) and FRAP \cite{AT-DC-trends}, 
has allowed molecular behavior to be deciphered on a short time-scale (msec). 
Therefore, it is now 
possible to propose new explanations of how the stability and plasticity of synapses can be 
{accounted for by}
interactions between molecules present in various 
compartments 
such as the 
extracellular protein domains in the presynaptic membrane, 
 the plasma membrane (receptors and associated molecules), 
the cytosol (scaffold molecules) 
and extracellular matrix. 

The preferential and specific localizations of receptors at synapses
result from their interactions with sub-membrane scaffold proteins. 
Comparison with the neuro-muscular junction encouraged the postulate that scaffold proteins are involved
in the so-called {\it stabilization} and {\it increased density} 
 of the receptors at
synapses   \cite{AT-DC-trends}.
 These two concepts, often unduly mixed, 
were extended to most central synapses and believed to be the heart of synapse-specific receptor 
localization. 
 This was reinforced by the discovery and characterization of numerous scaffold 
molecules interacting with inhibitory \cite{Moss01} {}
or excitatory receptors \cite{ShengSala01}{}. 
These structural and biochemical 
observations have perpetuated the %
 notion  that 
 at steady state 
 receptors are fixed at synapses and 
that this accounts for their density. 
Although electrophysiology has long since provided evidence 
for the existence of extrasynaptic receptors\cite{Faber85},
 they were often thought to constitute a pool distinct from synaptic receptors. 
 More importantly, their physiological roles have been limited to activation by spillover of neurotransmitter outside the synaptic cleft during massive release 
 \cite{Kullmann98,Clark02,Momiyama03, kullmann2004} 
or during glutamate release by neighboring glia \cite{Newmann03}. 
The notion that extrasynaptic and 
synaptic receptors are separate entities was reinforced by the fact that some receptor isoforms have 
specific sub-cellular distributions.

Interactions between pre- and post-synaptic elements are also important in determining 
not only the localization of synaptic contacts but also their excitatory or inhibitory nature 
 \cite{ChihScience05,Sheng05}.
The key molecules in this ``balancing act'' are {postsynaptic} neuroligins,
which interact with the $\beta$-neurexins,
which are 
themselves located in the  presynaptic release active zone. 
On the postsynaptic side, 
they are likely to bind to scaffold proteins. 
Therefore, { the postsynaptic } neuroligins provide the localization 
signal for the specific accumulation of given receptors at inhibitory or excitatory synapses. 
Without 
entering into detail, one of the most interesting features of this system is that these molecules, 
which induce either excitatory or inhibitory synapses, underpin the control of
excitation-inhibition balance. 
Other adhesive molecules such as N-cadherins are 
involved  in the homomeric interaction linking the presynaptic and postsynaptic membranes.

\begin{figure}[h]
\centering
\subfigure[\null] 
{   \label{fig:1}
  \includegraphics[width=7cm]{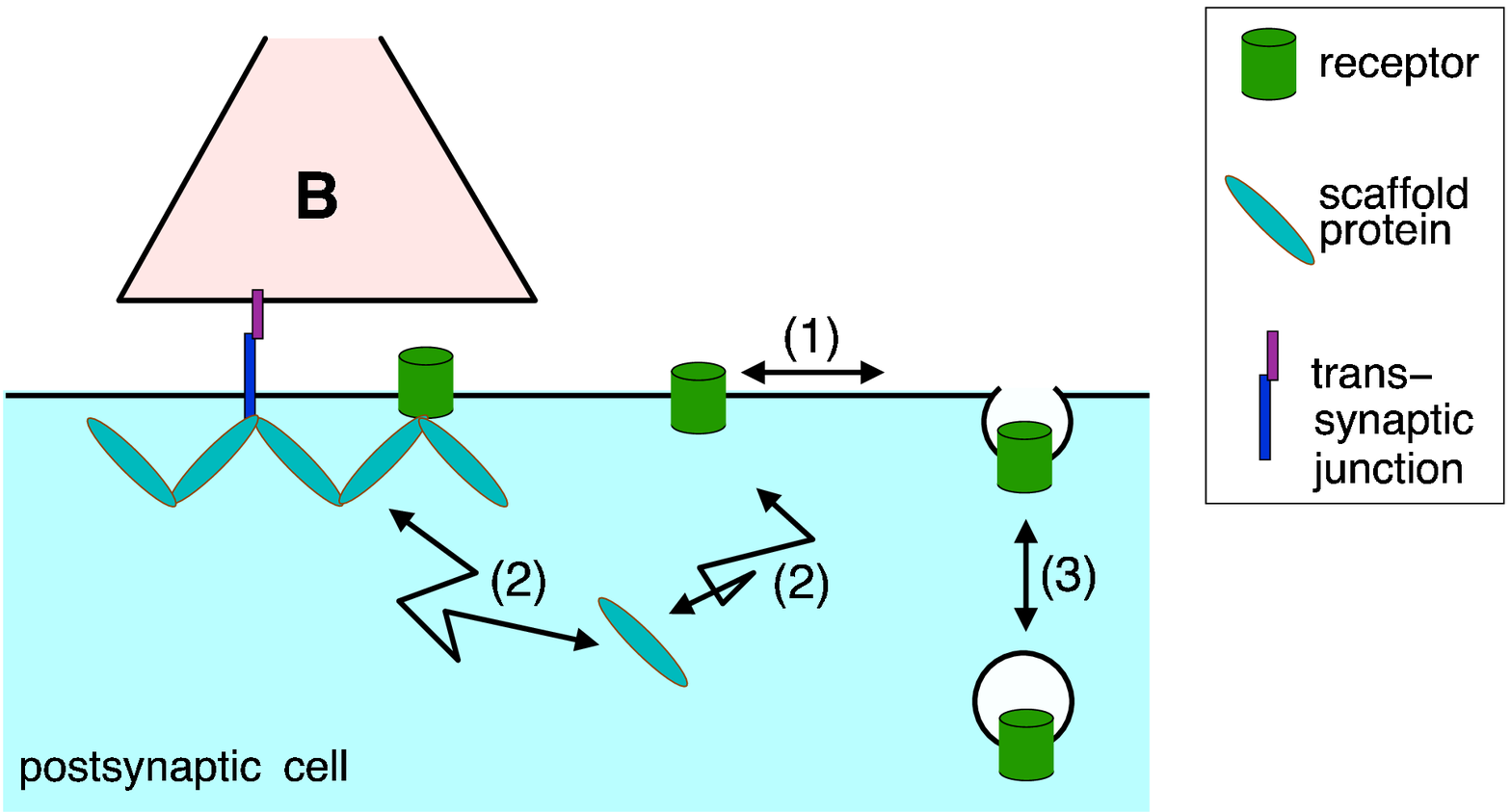}}
\hspace{1cm}
\subfigure[\null] 
{   \label{fig:figS1b}  \includegraphics[width=7.5cm]{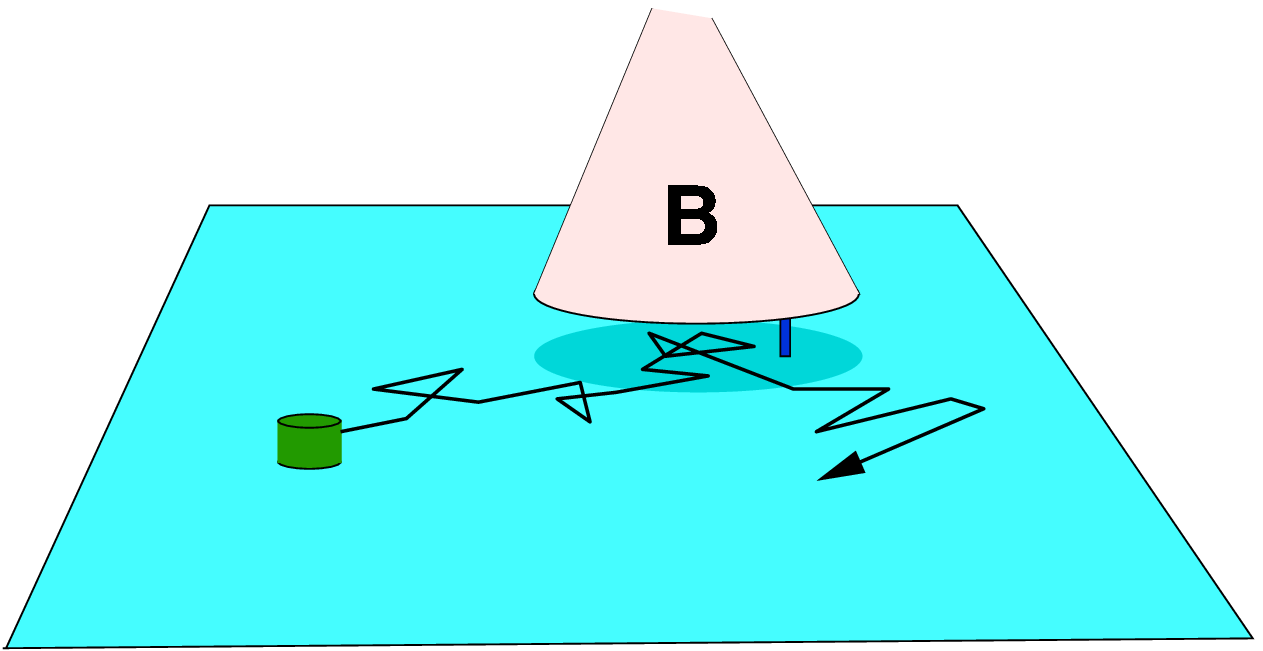}}
\caption{%
(Color online)\,
(a) 
Generic description of molecular mechanisms involved in the accumulation of receptors in front 
of terminal buttons (B). The arrow indicates: (1) the membrane diffusion of receptors; (2) the cytoplasmic 
diffusion of scaffold proteins and their binding to receptors; and (3) the
endocytosis/exocytosis of receptors. %
\\
(b) Schematic representation of the diffusive motion of receptors at the cell
surface   \cite{DC-AT}. %
}
\label{fig:diff-equiv} 
\end{figure}
The generic organization of the synapse is given in
Fig.~\ref{fig:diff-equiv}.
Receptors are indirectly linked to the presynaptic terminal buttons {\it via} scaffold proteins 
and trans-synaptic { homophilic or} heterophilic molecular interactions. 
We seek to link this topological organization to the movements of
both receptors and scaffold proteins. 
This minimal picture 
 holds for both excitatory and inhibitory 
synapses. 
There are more species of receptors and scaffold proteins at a given synapse than shown in the figure, and the molecular organization can be rather complex. 
In this study we have homogenized synaptic structure to account for diffusing receptors and scaffold proteins as a global entity. 
In fact the dynamic and static aspects of a system can be viewed differently depending on the resolution 
of experimental observation or 
model description. 
At the molecular level, thermal agitations 
cause both the spatial Brownian motion 
and chemical fluctuations of constituent molecules, 
added to which are the driving forces due to interaction among the molecules. 
On a mesoscopic level, the molecules are observable exclusively through their densities, and the thermal agitations 
are perceptible only as diffusion. 
Therefore, once the diffusion has reached a stationary or 
quasi-stationary state, the stability of a spatial density profile on a
mesoscopic level can {\it coexist} with 
the microscopic fluctuations of constituent molecules mentioned above. 
{ This fact, which was recognized in the late 19th century in the context of 
gas kinetics, can be applied to many other problems where we discuss a phenomenon on two
different scales.}
In particular, there are cases where the stationary state can be achieved with negligible net
fluxes of energy 
and material species, a situation called  {\it quasi-equilibrium}.
Such situations are characterized by 
the balance of chemical potentials of molecules both in space and in the chemical species in which the 
molecules move around. 
The peculiarity of the (quasi-)equilibrium state compared with other 
steady states is that the balance conditions of chemical potentials,
called the detailed balance condition in statistical physics, contain 
{\it no} kinetic parameters \cite{landau-std}.
In the present paper, we explore
a mesoscopic description of the quasi-equilibrium in the postsynaptic molecular
architecture. 
{ The rationale and consequences of the model are explained in general terms
  more accessible to biologists in Appendix C.}
The complexity of the synapse can in fact be accounted for by 
extending the number of zones and layers, as will be defined
in  Fig.~\ref{fig:threelayers}.
{We neglect the interaction between scaffold proteins and actin
  cytoskeletons (see also \S~\ref{subsec:2B}). It has been shown
  that the
 postsynaptic scaffolds of excitatory and inhibitory synapses in hippocampal neurons maintain their core components independent of actin filaments and microtubules.
 \cite{Allison00}.}


\section{Physical interpretation} \label{sec:kinetics}

\subsection{Reciprocal stabilization}  
The general picture that we propose in the present paper is that
receptors accumulating in 
front of the presynaptic release site are ``stabilized'' by scaffolding
molecules. 
The locus of the synaptic contact is supposed to be ``determined'' by
homophilic or heterophilic interactions between the pre- and  post-synaptic
membranes. 
{The stabilizing mechanism of the receptor density through the interaction 
with sub-membrane substances has also been explored in the context of 
cell adhesion\cite{Roux05,Smith-Udo} or of cellular recognition\cite{Pincet04},
or the polymer adsorption by surfactants\cite{AndelmanJF}.
A distinct feature of the present case of synaptic assembly is its reciprocal nature:
 The sub-membrane substances (scaffold proteins) are also assembled by the
 molecules on the membrane (receptors), while in the former cases it was large
 objects like colloids \cite{Roux05}, vesicles \cite{Smith-Udo}, micron-size
 particles \cite{Pincet04} or polymers \cite{AndelmanJF} that interact with many 
 molecules on the membrane. }

\subsection{Decoupling of kinetics from energetics in quasi-equilibrium}
\label{subsec:2A}

\begin{figure}[h] \includegraphics[scale=0.4,] 
{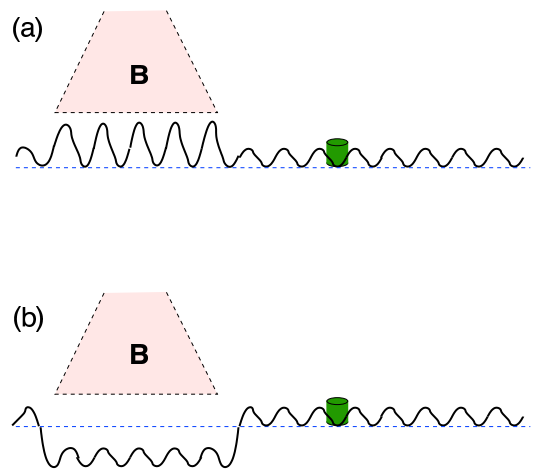} 
\caption{(Color online) 
 Kinetic and energetic components involved in receptor mobility and accumulation; (a,b) Potential 
energy profile (thick wavy lines) for a receptor (green object). Note its alterations below the presynaptic 
bouton  ({\bf {\sf B}}), illustrating two extreme situations. Compared to extrasynaptic membrane, the energy barrier 
can be higher (a) or the energy level lower (b). The consequences are that (a) the diffusion is slowed 
down beneath the synaptic bouton but the density of receptors can be identical at synaptic and extrasynaptic 
membrane in the steady state; (b) that the diffusion coefficients can be identical within the two zones but receptor density is higher beneath the synaptic bouton. Experimental data (accumulation of receptors and lower 
diffusion coefficient) \cite{DC-AT}  indicate that a combination of the two is
responsible for accumulation  of receptors. 
}
\label{fig:DiffDens}  
\end{figure}

In the context of the problem and the minimal model of quasi-equilibrium presented above, we 
will briefly describe the {\it separation} of kinetic aspects from static ones
mentioned in the introduction  (see Fig.~\ref{fig:DiffDens}).  
{
The conclusion is that, in the quasi-equilibrium situation, the
accumulation of receptor density under the synapse should 
not be ascribed to kinetic mechanisms such as 
small mobility of receptors inside a synaptic zone,
but to the static aspect of molecular interactions.
}

Fig.~\ref{fig:DiffDens} (a) shows a potential profile for a receptor diffusing on the membrane with higher barriers 
inside than outside synapses. 
Obstacles within synapses create potential barriers which modify the 
kinetics (reduced diffusion), but do not necessarily create higher receptor density at steady state. 
{ One can show by a simple calculation that 
if the rightward and leftward transition rates across each barrier are
symmetric, the probability of finding the receptor is homogeneously distributed
in the steady state.} 
By contrast, in  Fig.~\ref{fig:DiffDens}(b) 
the mean level of the potential valley is lowered within synapses, but the potential barriers are 
unchanged there. 
As a consequence, with potential barriers of the same height inside 
and outside synapses, receptors diffuse equally fast in extrasynaptic and synaptic regions, although 
the density is increased in the latter. 
This simplistic schematic representation again emphasizes  that 
postsynaptic accumulation and diffusivity are two independent physical
characteristics. 
{We note that, as this separation is strictly valid only at equilibrium, 
it is not a mere temporal analogue of the concept of the
  compatibility between microscopic fluctuations and mesoscopic steady state 
mentioned in the previous section.
 Below we will identify the time-window where we can apply  
approximately the theoretical framework of quasi-equilibrium 
to the processes of receptors and scaffold proteins.}

\subsection{ Summary of time scales and justification of quasi-equilibrium treatment}
\label{subsec:2B}

\begin{figure}[h] \includegraphics[scale=0.6,] 
{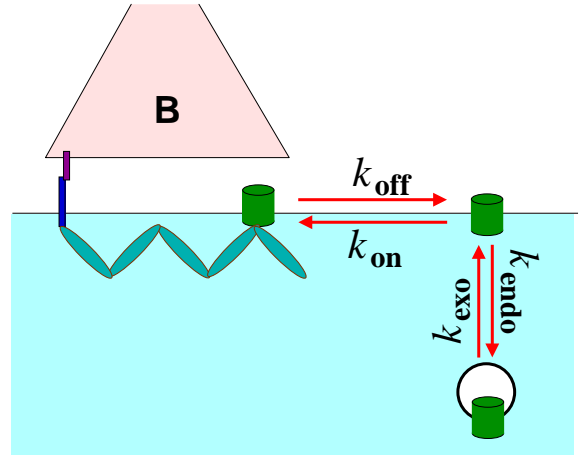}
 \caption{(Color online)
Kinetic parameters and cellular biology of receptors. Exo/endocytosis and synaptic to extrasynaptic 
transfer are characterized by specific rate constants, $k_{\rm endo/exo}\ll k_{\rm on/off}$. 
The half-life of receptors in the plasma membrane 
(on the order of tens of minutes to half a day) and the dwell time of receptors at synapses 
are given in terms of these parameters.
}
\label{fig:figS1bis}
\end{figure}

Neurotransmitter receptors 
 undergo both lateral diffusion on the plasma membrane and cycling
through exo-/endo-cytosis between the plasma membrane and cytoplasmic vesicles. 
We postulate two characteristic time scales: 
\bde 
\item{$\tau_{\rm R, eq}$ :} (quasi-)equilibration time of the receptors on the postsynaptic cell membrane
\item{$\tau_{\rm R, cyc}$ :} recycling time of receptors related to  endocytosis and exocytosis
\ede
The rate of receptor exchanges
 $k_{\rm on}$ and  $k_{\rm off},$ $k_{\rm endo}$ and  $k_{\rm exo}$ 
 allows the computation of 
 $\tau_{\rm R, eq}=(k_{\rm on}+k_{\rm off})^{-1}$ and 
 $\tau_{\rm R, cyc}=(k_{\rm endo}+k_{\rm exo})^{-1}$, respectively
 (Fig.~\ref{fig:figS1bis}).
Experimental evidence indicates  that $\tau_{\rm R, eq}$ ranges
from tens of seconds to minutes, and 
 $\tau_{\rm R, cyc}$ ranges from  tens of minutes to about half a day   \cite{Derkach07,Rasmussen02}.
The scaffold proteins also experience movements between the plasma membrane periphery 
and the bulk cytoplasm. 
Furthermore, local amounts of scaffold proteins in the bulk cytoplasm 
are regulated by means of expression/degradation, or by transport-associated compartmentalization. 
Here again, we can postulate two characteristic time-scales: 
\bde 
\item{ $\tau_{\rm s, eq}$ :} (quasi-)equilibration time for the migration of scaffold proteins 
\item{ $\tau_{\rm s, cyc}$ :} recycling time related to the synthesis and degradation of scaffold proteins
\ede
Experimental evidence indicates that 
$\tau_{\rm s, eq}$ is of the order of minutes to tens of minutes \cite{temps1,temps2}, 
while $\tau_{\rm s, cyc}$ is likely to be several hours. %

We can thus estimate the time window for quasi-equilibrium as between minutes and hours. 
That is, when (i) both the number of receptors on the plasma membrane and the density of scaffold 
proteins in the cytoplasm remain almost constant, 
while (ii) the membrane diffusion of receptors 
and the cytoplasmic diffusion of scaffold proteins have reached equilibrium. 
We therefore focus on the time window 
 $\Delta t$ for observation/description with the following limits:
\begin{equation}
\max\{\tau_{\rm R, eq},\tau_{\rm s, eq}\}
\lesssim
\Delta t
\lesssim
\min\{\tau_{\rm R, cyc},\tau_{\rm s, cyc}\}.
\end{equation} 
and develop in the following section
 a quasi-equilibrium model using assumptions (i) and (ii).

One might ask if the actin cytoskeleton forms a network underneath the
scaffold proteins and works as a frozen heterogeneous background. 
The recent FRAP analyses, however, have shown that,
about 85 \% of actins in dendritic spines are turned over within 44 seconds\cite{Star02},
and also the turnover of $\alpha$-actinin (passive actin-binding protein) is
 more rapid than  that of PSD-95, a scaffold protein of the excitatory 
synapse \cite{temps2}.
 Therefore, within the time window $\Delta t$ defined above, we assume that
 the actin 
cytoskeleton is a fluid-like background and ignore it in our minimal model.

\section{Mesoscopic model and phase-equilibria} 
\label{sec:model}
\subsection{Spatial compartments and density variables}
\label{subsec:param}

\begin{figure}
\centering
\subfigure[] 
{    \label{fig:sub:a}  \includegraphics[scale=0.45,] 
{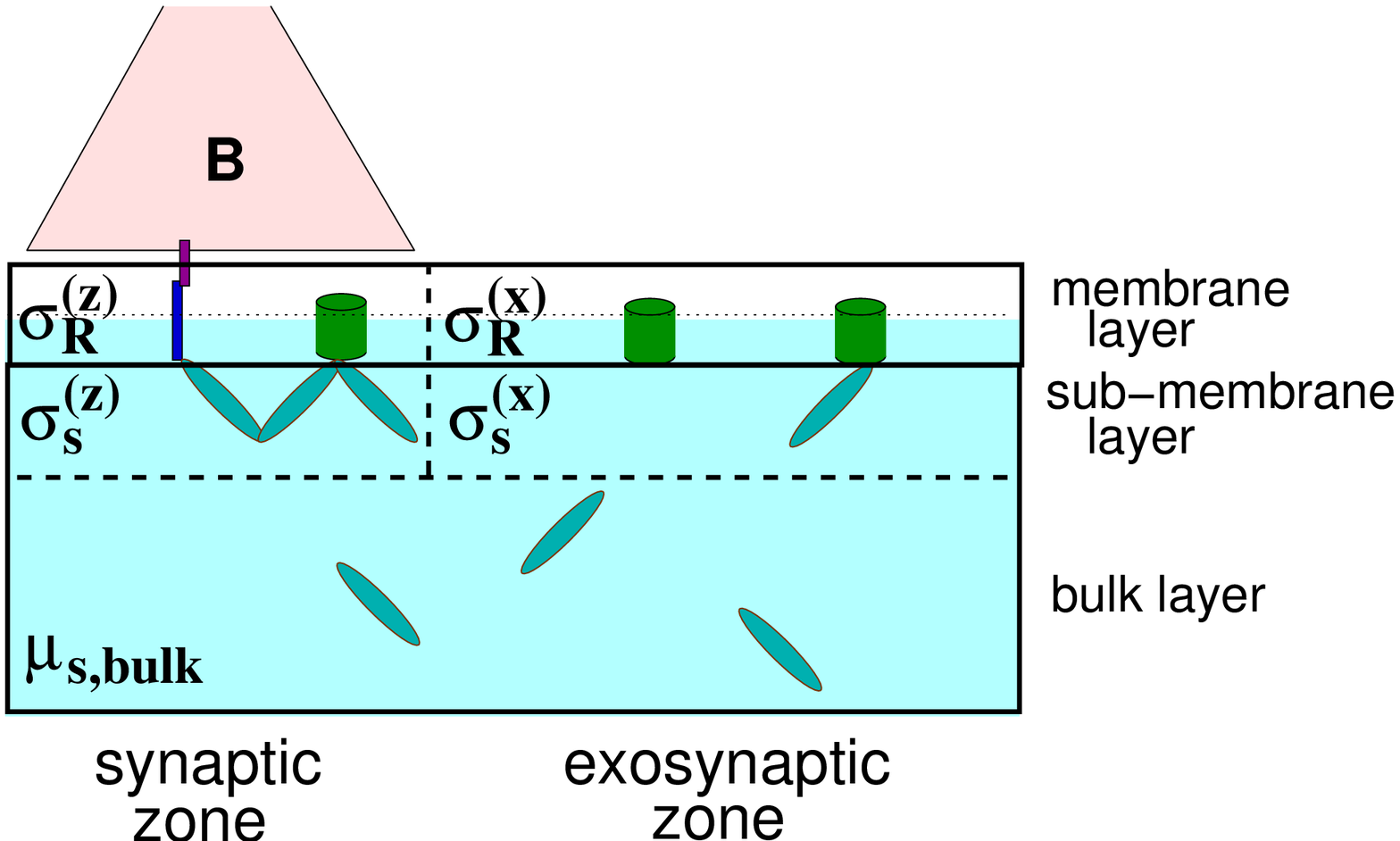} }
\hspace{0cm}
\subfigure[] 
{    \label{fig:sub:b}  \includegraphics[scale=0.45,] 
{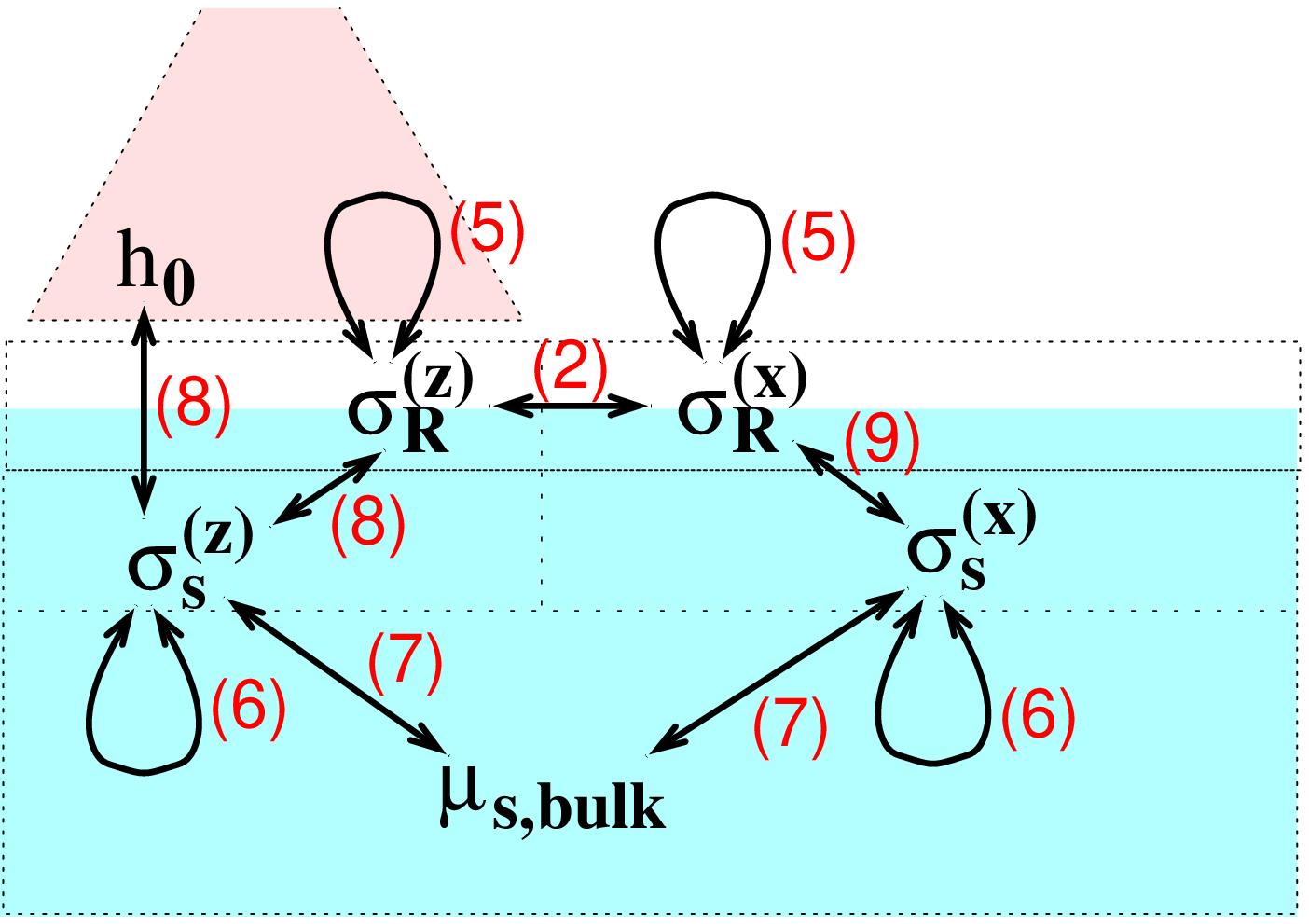}}
\caption{(Color online) 
{(a)} 
Three-layer, two-zone model: The model assumes a three-layer partition of the postsynaptic 
cell: {\it membrane layer}, {\it sub-membrane layer} and {\it bulk layer}.
In the first two layers, we establish a spatial 
partition with two zones: a synaptic (z) and an extrasynaptic (x) zone, where the areal densities ($\sigma$) are 
used as variables, and receptors ({\bf R}) and scaffold proteins ({\bf s}) are indicated as suffixes. 
In the bulk layer, the 
density of scaffold proteins corresponds to chemical potential, $\mu_{\rm s, bulk}$. 
The receptors can diffuse within the 
membrane layer, and the scaffold proteins diffuse among the zones in both the sub-membrane layer and the 
bulk layer. \\
(b) Correlations among molecules: The arrows indicate the molecular correlations taken into 
account in the present model. The numbers like (2) etc. correspond to those of equations in the text. 
}
\label{fig:threelayers} 
\end{figure}
%
The quasi-equilibrium defined above will be assumed in the homogenized 
schema of the post-synaptic cell 
 (Fig.\ref{fig:threelayers} (a)). 
We assume three layers along the vertical direction to the membrane:
The outmost layer is the {\it membrane layer} with all the receptors and other
trans-membrane signaling proteins (see below).
The intermediate and {\it sub-membrane layer} (a few nanometers) constitute the cytoplasmic volume where scaffold proteins interact with receptors and other 
trans-membrane molecules (e.g. adhesion molecules). 
The innermost layer is the  {\it bulk cytoplasm}, which is  the reservoir 
of scaffold proteins that swap with the sub-membrane layer.

Laterally,  we define   {\it synaptic} (superscript: z) and  
{\it extrasynaptic} (superscript: x) zones.
This 
partition can be justified since the time-scale of modeling is greater than the equilibration time 
of both receptors and scaffold proteins. 
However, we have neglected possible mesoscopic 
substructures within the synaptic zone, a point to be considered in future investigations. 
The 
reservoir of scaffold proteins is common to these two zones. 
Within these five compartments, we attribute densities to membrane receptors
and sub-membrane scaffold proteins as follows, see
Fig.\ref{fig:threelayers}(a). 
\begin{description}
\item{ $\sigma^{\rm (z)}_{\rm R}$  and $\sigma^{\rm (x)}_{\rm R}$ :}
number of receptors (suffix: R) per surface area (areal density)
  in the synaptic  
and extrasynaptic zones, respectively,
\item{$\sigma^{\rm (z)}_{\rm s}$  and $\sigma^{\rm (x)}_{\rm s}$ :}
number of scaffold proteins (suffix: s)  per surface area (areal density) 
 in the sub-membrane synaptic  and extrasynaptic  zones, respectively.
\end{description}
Here superscripts $^{\rm (z)}$ [$^{\rm (x)}$] denote
the quantities associated with the synaptic zone [extrasynaptic zone],
respectively.

The total number of receptors on the membrane is constant within the time-scale of modeling, and 
is expressed by 
\begin{equation}\label{eq:NtotR}
  N_{\rm R}=A^{\rm (z)} \sigma^{\rm (z)}_{\rm R} %
+ A^{\rm (x)} \sigma^{\rm (x)}_{\rm R} %
=\mbox{constant},
\end{equation}
where  $N_{\rm R}$ is the total number of membrane  receptors, and
 $A^{\rm (z)}$ and $A^{\rm (x)}$ are the surface areas of the synaptic zone and  extrasynaptic zone,
respectively. 
{Experimental data indicate that receptors can be exchanged between 
synaptic sites 
 \cite{DC-AT} and, therefore, the membrane can be considered as a global field where synaptic
contact introduces a singularity allowing for the local accumulation of the constituent elements of the postsynaptic machinery. 
Thus, each synapse behaves as a donor or acceptor of molecules. }

As for the mechanism determining the spatial extension of the 
PSD, one might consider a physical mechanism
which minimizes the free energies due to surface (peripheral) 
contribution and the bulk (areal) contribution. 
{
A possible origin is
  entropic, that is, the steric repulsion among molecules reflecting their
  three-dimensional geometrical arrangement.
Such} situation is well exemplified in recent work on syntaxin 1 clusters \cite{Size-Lang-Sci07}.
However, the actual size of the synaptic density matches 
 the size of the presynaptic active zone. 
We therefore will not elaborate on this issue and simply assume 
here that the size of PSD is determined externally.
The size of the PSD is likely to be correlated with the number of scaffold proteins.  
The total number of scaffold proteins in the sub-membrane layer 
can fluctuate despite a constant density in the layer of bulk cytoplasm.
As a consequence,, scaffold protein chemical potential is an important parameter (see 
the text below and \citer{eq:Gbulk}).
 
\subsection{Construction of free energy}
\label{subsec:free-energy}
The observed densities of constituent molecules in 
the quasi-equilibrium state correspond to the maximum probability of
realization.
Following Gibbs' statistical mechanics, this  
probability is given by the Boltzmann factor,  $e^{-G/\kT}$,
where $G$ is a pertinent (Gibbs) free energy function for the whole system. 
The maximum of this factor defines the (Boltzmann) equilibrium.
$G$ is the sum of the contributions from each compartment,
\beq \label{eq:G}
G=A^{\rm (z)} g^{\rm (z)}+ A^{\rm (x)} g^{\rm (x)},
\eeq
where $g^{\rm (z)}$ [$g^{\rm (x)}$] are the {\it free energies} per unit area of the membrane in the synaptic [extrasynaptic] zone, respectively. 
The variables of these free energies will be introduced below. 
Experimental data suggest that, in our minimal model, 
 $g^{(\alpha)}$ ($\alpha=z$ or $x$)
can be constructed from the following components :
\beq \label{eq:Glayers}
g^{(\alpha)}=g^{(\alpha)}_{\rm mem}+g^{(\alpha)}_{\rm sub}
+g^{(\alpha)}_{\rm bulk}+g^{(\alpha)}_{\rm mem-sub}.
\eeq
Here the first three terms denote the contributions from each layer, i.e. the membrane layer (mem), 
sub-membrane layer (sub) and bulk layer (bulk), respectively, and the last term is the key term representing the interactions between the first two layers.
 The biological counterparts of 
$g^{(\alpha)}_{\rm mem}$, $g^{(\alpha)}_{\rm sub}$,
$g^{(\alpha)}_{\rm bulk}$ and $g^{(\alpha)}_{\rm mem-sub}$
correspond to the free energy associated with receptors in the plasma
membrane ($g^{(\alpha)}_{\rm mem}$), 
scaffold proteins in the sub-membrane layer (i.e.,  
scaffold proteins in the bulk cytoplasm in relation 
to specific domains)($g^{(\alpha)}_{\rm sub}$), 
scaffold proteins in the rest of the bulk cytoplasm ($g^{(\alpha)}_{\rm
  bulk}$),  
and scaffold-transmembrane protein interactions ($g^{(\alpha)}_{\rm mem-sub}$),
respectively.  
We detail these terms below
(see also Fig.~\ref{fig:threelayers}(b)).

\begin{figure}[h]
  \includegraphics[scale=0.8,] {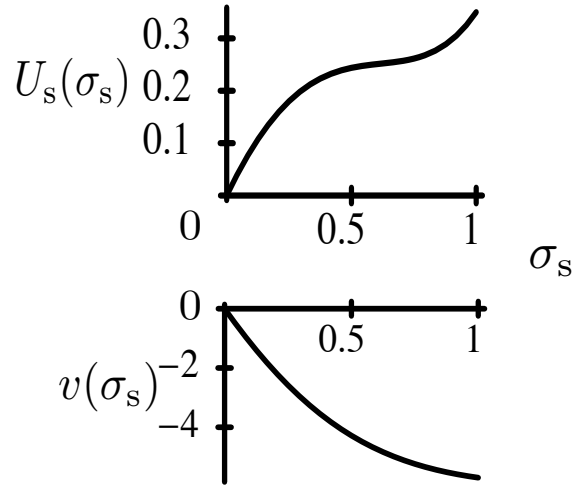} %
\caption{$U_{\rm s}(\sigma_{\rm s})$ {\it vs}  $\sigma_{\rm s}$  (top) and 
 $v(\sigma_{\rm s})$ {\it vs} $\sigma_{\rm s}$ (bottom).
}
\label{fig:Us} 
\end{figure}

\bde
\item{\underline{Membrane layer} } :\\
The term $g^{(\alpha)}_{\rm mem}$ contains 
the density of the receptors in the corresponding zone,
$\sigma^{(\alpha)}_{\rm R}$, and we assume no direct binding  interaction between  receptors
{\it except for} the lateral steric exclusion: 
\beq \label{eq:Gmem}
g^{(\alpha)}_{\rm mem}(\sigma^{(\alpha)}_{\rm R})
= \kT \left[\sigma^{(\alpha)}_{\rm R} \log
\frac{\sigma^{(\alpha)}_{\rm R}}{\sigma_{{\rm R}0}}  
+(\sigma_{{\rm R}0}-\sigma^{(\alpha)}_{\rm R}) \log
\frac{\sigma_{{\rm R}0}-\sigma^{(\alpha)}_{\rm R}}{\sigma_{{\rm R}0}}\right],
\eeq
where $\sigma_{{\rm R}0}$ is the saturation density, which we assumed 
to be common to the two zones.
\citer{eq:Gmem} was deduced from the factor  
$e^{-(A^{(z)}g^{(z)}_{\rm mem}+A^{(x)}g^{(x)}_{\rm mem})/\kT}$,
which gives the combinatorial number for spatial distribution 
of the receptors on the membrane.  
This equation \citer{eq:Gmem} establishes the relationship between the geometrical distribution of individual receptors and the (free) energy of a collection of receptors. 
\item{\underline{Sub-membrane layer} } :\\
The term $g^{(\alpha)}_{\rm sub}$, which has the same form as 
in \citer{eq:Gmem}, accounts for scaffold proteins. 
In addition to geometrical volume exclusion,this equation takes into account a specific attractive 
interaction among scaffold proteins 
($U(\sigma^{(\alpha)}_{\rm s})$).
\beq \label{eq:Gsub}
g^{(\alpha)}_{\rm sub}(\sigma^{(\alpha)}_{\rm s})
= \kT \left[\sigma^{(\alpha)}_{\rm s} \log
\frac{\sigma^{(\alpha)}_{\rm s}}{\sigma_{{\rm s}0}}  
+(\sigma_{{\rm s}0}-\sigma^{(\alpha)}_{\rm s}) \log
\frac{\sigma_{{\rm s}0}-\sigma^{(\alpha)}_{\rm s}}{\sigma_{{\rm s}0}}\right]
+U(\sigma^{(\alpha)}_{\rm s}) ,
\eeq
where $\sigma_{{\rm s}0}$ is the saturation (areal) density of the scaffold proteins.
The last term $U_{\rm s}(\sigma^{(\alpha)}_{\rm s})$ representing the 
non-combinatorial part of the free energy
includes  the entropic cost of confinement 
 ($U_1$), the mutual attraction among the 
scaffold proteins ($U_2$) and the specific saturation effect among them
 ($U_4$),
which imposes a smaller limiting value than $\sigma_{{\rm s}0}$.
Recent molecular studies \cite{Bedet-JBC06} on the scaffold
  protein for the inhibitory synapse (gephyrin) have identified 
trimerization and dimerization domains.
They may be responsible for the hexagonal oligomerization of the postsynaptic
scaffold organization \cite{gephyerin-TINS08}. 
The attraction by $U_2(<0)$ and non-steric saturation $U_4(>0)$ reflects these
findings.
We therefore propose for $U_{\rm s}(\sigma^{(\alpha)}_{\rm s})$ the
following function: 
$U_{\rm s}(\sigma_{\rm s})
=U_1 \sigma_{\rm s}+U_2 {\sigma_{\rm s}}^2
+U_4 {\sigma_{\rm s}}^4,$ 
with the coefficients $U_1>0$,  $U_2<0$ and  $U_4>0$. 
see Fig.~\ref{fig:Us}(top).
The most important term is $U_2$ (the attractive term)
because  $U_1$ can be included as a shift of the chemical potential
of the reservoir (see below), while the last term
 $U_4$ acts effectively %
as steric repulsion.
 
\item{\underline{Bulk layer}} :\\
The term $g^{(\alpha)}_{\rm bulk}$ represents the free energy associated with
scaffold proteins of the bulk cytoplasm. It is characterized only by the
chemical potential of these scaffold proteins, which we denote by $\mu_{\rm s,bulk}$. 
Although there is a single contribution to $G$ from the 
scaffold proteins in the bulk cytoplasm, $(-\mu_{\rm s,bulk}) (A^{\rm (z)}\sigma^{(z)}_{\rm s}
+A^{\rm (x)}\sigma^{(x)}_{\rm s}),$ it can be separated 
in two parts, i.e. $A^{\rm (z)}g^{(z)}_{\rm bulk}$ and 
$A^{\rm (x)}g^{(x)}_{\rm bulk}$,
linked to synaptic (z) and extrasynaptic (x) zones, respectively: 
\beq \label{eq:Gbulk}
g^{(\alpha)}_{\rm bulk}(\sigma^{(\alpha)}_{\rm s})
=-\mu_{\rm s,bulk}\,\sigma^{(\alpha)}_{\rm s}.
\eeq
In biological terms, an increase in scaffold proteins in the bulk cytoplasm will increase 
 $\mu_{\rm s,bulk}$,
and therefore the capacity of these proteins to be involved  in the clustering of postsynaptic 
receptors.

\item{\underline{Membrane/sub-membrane interface}} :\\
The formal description of the interactions between compartments must take into account their interfaces. 
The interface 
  for molecular interactions sets  
 a discontinuity in the molecular organization
of the synapse.
Depending on the zone, the interaction free energy, $g^{(\alpha)}_{\rm mem-sub}$ contains one or two contributions: 
The interaction between membrane receptor and scaffold protein, and additional interaction between scaffold protein and a trans-membrane protein involved in pre-to-postsynaptic signaling for the localization of the contact. 
The latter 
contribution is denominated $h_{0}$, 
 and behaves as an attracting field 
 (See Fig.~2(b)), introducing a local energetic component recruiting scaffold proteins. 
The interaction free energy 
at the synapse is now expressed as: %
\beq \label{eq:memsubA}
g^{(z)}_{\rm mem-sub}
(\sigma^{(z)}_{\rm R},\sigma^{(z)}_{\rm s},h_{0})
= \sigma^{(z)}_{\rm R} \,\, v(\sigma^{(z)}_{\rm s})
-h_{0}\sigma^{(z)}_{\rm s} 
\eeq
and outside of synapse as:
\beq \label{eq:memsubB}
g^{(x)}_{\rm mem-sub}
(\sigma^{(x)}_{\rm R},\sigma^{(x)}_{\rm s}%
)
= \sigma^{(x)}_{\rm R} \,\, v(\sigma^{(x)}_{\rm s}). 
\eeq
The first term in both equations represents the interaction between membrane receptor 
and scaffold protein, and depends on receptor and scaffold protein density. 
$v(\sigma^{(x)}_{\rm s})$ should reflect (i) linearity in the dilute regime, (ii) curvature for 
intermediate regime, and (iii) saturation at high concentration regime.
The saturation is related 
to the steric hindrance of molecules and to the number of binding sites available on a receptor for interaction with scaffold proteins. 
{We have tried the following two forms: 
 {\it (1)} $v(\sigma_{\rm s})
=v_{\rm f}[1-e^{
-v_1(\sigma_{\rm s}/\sigma_{{\rm s}0})
-v_2(\sigma_{\rm s}/\sigma_{{\rm s}0})^2}]$ 
(see Fig.\ref{fig:Us} (bottom)) and 
{\it (2)} $v(\sigma_{\rm s})
=v_{\rm f}[v_1(\sigma_{\rm s}/\sigma_{{\rm s}0})
-\tilde{v}_2(\sigma_{\rm s}/\sigma_{{\rm s}0})^2],$
where  $v_{\rm f}(<0)$ corresponds to 
the specific attractive power between the two group of molecules, while
 $v_1(>0)$ and $v_2(>0)$ or $\tilde{v}_2(>0)$ realize the above three features, (i)-(iii). 
The overall characteristics of $v(\sigma_{\rm s})$ in {\it (2)} are
similar to Fig.\ref{fig:Us} (bottom) for $0<\sigma_{\rm s}/\sigma_{{\rm s}0}<1$.
 It turns out that  the qualitative results of the numerical analyses
 are robust against the choice between the types
 {\it (1)} and {\it (2)}, and we will present below the results for case
 {\it (1)} only.} 
That we have retained only the linear dependency 
on $\sigma^{(x)}_{\rm R}$ is based on the observation
 that the number of receptors at a synaptic site is {usually}
well below the stoichiometric limit determined by the number of underlying
scaffold  proteins.  
The number of receptors  present in a PSD
 is below 100 for excitatory \cite{Sheng-AnnuRev07} 
and  inhibitory \cite{shigemoto07} synapses.
In contrast, the number of scaffolding molecules such as PSD-95 in
excitatory postsynaptic differentiations is  about 300 \cite{PSdensity-Sheng}.
Therefore, the ratio of receptor to scaffold binding sites is likely to be below  50\%.
The second term of \citer{eq:memsubA} represents the positive bias for
  the  
scaffolding molecules due to the transsynaptic signal, and therefore exists
only in the synaptic zone.
This signal is carried through the interaction between the transmembrane
molecules. 
The range of $h_0$ is such that this bias is reversible and does not exceed too
much the  order of $\kT$.
\ede 
In biological terms, the expression of the free energies for the membrane/sub-membrane interface 
accounts for the network of molecular interactions between presynaptic terminals through adhesion 
 ($h_0$),  scaffold proteins
  ($\sigma^{(z)}_{\rm s}$) and receptors   ($\sigma^{(z)}_{\rm R}$). 
  This will now allow us to sum the contributions from the layers and their interfaces to obtain the free energy 
 $G$, which will be used in the 
next section to establish the conditions of the quasi-equilibrium. 

\subsection{Phase equilibria}
\label{subsec:chemical-potential}
What we will denominate below as the {\it phase} is any realization of physical
  states that corresponds to the minimum of the model free energy function
  (``Landau function'') with respect to its variables specifying
physical states. In the present model the variables are the densities,
$\{\sigma^{\rm(z)}_{\rm R},\sigma^{\rm(x)}_{\rm R},\sigma^{\rm(z)}_{\rm s}, 
\sigma^{\rm(x)}_{\rm s} \}$.
In this case, a phase can represent spatially heterogeneous distributions
of membrane receptors and sub-membrane scaffold proteins.
The {\it phase change} is then the phenomenon where the distribution of these
molecules changes in a discontinuous manner as some model parameters are
changed continuously across a transition point. %

{The phase change can be strictly defined and realized only if
the system that  a model represents is infinitely large.
Otherwise, the thermal fluctuations in the vicinity of the transition 
point may cause the temporal switching between one phase to the other.
Therefore, characteristic switching time depends on the system size.
The present model deals with synaptic buttons, which are on a mesoscopic
scale.
In each synaptic bouton the PSD contains receptors and scaffold proteins 
of the order of tens ($\sim$50 \cite{Signal-Kennedy}) and hundreds 
($\sim$300 \cite{PSdensity-Sheng}), respectively (see \cite{Sheng-AnnuRev07} and the references cited therein).
Apparently the lifetime of each PSD %
is  long so that its eventual dissolution, which 
corresponds to the switching from the localized phase to nonlocalized
phase (see below), is not observed, though it is {\it in principle} possible.
We, therefore, suppose that the thermodynamic framework  describing the phase
change is practically applicable to our system.}%

{As mentioned above the (quasi-)equilibrium states %
will be looked for in a space with four variables, 
$\{\sigma^{\rm(z)}_{\rm R},\sigma^{\rm(x)}_{\rm R},\sigma^{\rm(z)}_{\rm s}, 
\sigma^{\rm(x)}_{\rm s} \}$.
The Landau function in our model is $G$ (see (\ref{eq:G})),
which includes the free energies related to the interfaces between 
the compartments as  represented in Fig.~\ref{fig:threelayers}(a). 
The highest probability of realization corresponds to the 
maximum of $\propto e^{-G/\kT}$, or the minimum of $G$,
provided that the total number of membrane receptors is constrained 
to be constant, (\citer{eq:NtotR}). 
We use a standard technique of the Lagrange multiplier (see
Appendix~A.1 for a brief description), which replaces the problem of  
constrained optimization by the following conditions, }
${\partial}[G- \mu_{\rm R}^*
(A^{\rm (z)}\sigma^{\rm (z)}_{\rm R}+ A^{\rm (x)} \sigma^{\rm (x)}_{\rm R})
]/{\partial \sigma^{(\alpha)}_{\rm R}}=$
${\partial}[G- \mu_{\rm R}^*
(A^{\rm (z)}\sigma^{\rm (z)}_{\rm R}+ A^{\rm (x)} \sigma^{\rm (x)}_{\rm R})
]/{\partial \sigma^{(\alpha)}_{\rm s}}=0,$
for $\alpha=z$ and $x$, 
or, 
\beq  \label{eq:cond-eq}
\frac{\partial G}{\partial \sigma^{(z)}_{\rm R}}-\mu_{\rm R}^* A^{\rm (z)}
=
\frac{\partial G}{\partial \sigma^{(x)}_{\rm R}}-\mu_{\rm R}^* A^{\rm (x)} 
=\frac{\partial G}{\partial \sigma^{(z)}_{\rm s}}
=\frac{\partial G}{\partial \sigma^{(x)}_{\rm s}}
=0,
\eeq
where the {Lagrange multiplier} $\mu_{\rm R}^*$
has the meaning of the chemical potential of the membrane receptors.
It is to be determined so that the constraint of \citer{eq:NtotR} is
satisfied.  
These conditions, five in total including
\citer{eq:NtotR}, are sufficient to determine the five unknown variables, 
$\{\sigma^{\rm(z)}_{\rm R},\sigma^{\rm(x)}_{\rm R},\sigma^{\rm(z)}_{\rm s},
\sigma^{\rm(x)}_{\rm s}, \mu_{\rm R}^*\}$. 
This approach was chosen because the existence of reciprocal interactions prevents a straightforward estimation of receptor number as a function of scaffold or 
trans-membrane signal protein number only.

{
Though the treatment of the model is very general and based on the
  principles of statistical thermodynamics, the architecture of the model is
  developed on the basis of 
  the following details known %
  about the synaptic sites:  
the presence of the localization signal ($h_0$), 
{
interactions between scaffold proteins (nonlinearity of $U_{\rm
  s}(\sigma_{\rm s})$), especially the intermolecular attraction (i.e. the
term $U_2\,{\sigma_{\rm s}}^2$ with $U_2<0$)} and  
interaction between scaffold proteins and receptor
molecule ($\sigma_{\rm R}v(\sigma_{\rm  s})$). 
  }

\section{Results: Localization-delocalization transition} 
\label{sec:results}
We analyze how the local density of receptors at the synapse in  
the quasi-equilibrium states depends on 
 control parameters represented by the pre-to-postsynaptic signaling  ($h_0$)
as well as the chemical potential of cytoplasmic scaffold proteins ($\mu_{\rm s,bulk}$). 
One should keep in mind that, 
since the total synaptic and extrasynaptic
number of receptors, $N_{\rm R}$, is supposed to be constant within 
the time scale of our interest, the chemical potential of the receptors, $\mu_{\rm R}^{*}$,  is not a controllable 
parameter (unlike that of scaffold protein, $\mu_{\rm s,bulk}$), 
but is a part of the output of the quasi-equilibrium 
condition. This is why we did not study the variation  {\it vs}  $\mu_{\rm
  R}^{*}$. 
 \citer{eq:NtotR} and \citer{eq:cond-eq} can be solved numerically 
(see Appendix~A.2 for technical details). 

The values of the parameters were chosen 
to account for the possible experimental situations of the system. 
 They
include the proportion of membrane covered by synaptic contact, $A^{\rm
  (z)}/A^{\rm (x)}$, where  
we have taken $(A^{\rm (z)},A^{\rm (x)})=(0.1,0.9)$ except for in
\S~\ref{subsec:parametres} 
where 
$(A^{\rm (z)},A^{\rm (x)})=(0.01,0.99)$,  
 the non-steric part of the free energy of
scaffold proteins in the sub-membrane, 
$U_{\rm s}(\sigma_{\rm s})=U_1 \sigma_{\rm s}+U_2 {\sigma_{\rm s}}^2
+U_4 {\sigma_{\rm s}}^4,$ with $\{U_1,U_2,U_4\}=\{1,-1.15,0.5\}$,
 and the factor  in the scaffold protein-receptor interaction energies (see
 (\ref{eq:memsubB})),
$v(\sigma_{\rm s})=v_{\rm f}[1-e^{-v_1(\sigma_{\rm s}/\sigma_{{\rm s}0})
-v_2(\sigma_{\rm s}/\sigma_{{\rm s}0})^2}],$
with $\{v_{\rm f},v_1,v_2\}=\{-6,2,1\}$.
To check  the robustness  (see below Eq. (\ref{eq:memsubB})), we used 
$v(\sigma_{\rm s})
=v_{\rm f}[v_1(\sigma_{\rm s}/\sigma_{{\rm s}0})
-\tilde{v}_2(\sigma_{\rm s}/\sigma_{{\rm s}0})^2],$
with $\{v_{\rm f},\tilde{v}_1,\tilde{v}_2\}=\{-6,1.9,1\}$.
The units of energy and space are chosen such that 
$\kT=1$ and  
the saturation areal density of receptors on the membrane,
${\sigma_{{\rm R}0}},$ and 
that of scaffold proteins in the sub-membrane layer, 
 ${\sigma_{{\rm s}0}},$ are 1 in both zones.

For a certain range of parameters, $\{h_0,\mu_{\rm s,bulk}\}$,
 \citer{eq:NtotR} and
 \citer{eq:cond-eq} have multiple solutions. 
 When it happens, the solution chosen is the 
one with the minimum value of $G$, and therefore the maximum probability of
realization, $e^{-G/\kT}$. 
The phase change between different solutions %
corresponds to  
the standard criterion of the so-called {Maxwell's construction,}
which was originally used in the Van der Waals model of vapor-liquid
condensation (see below).

\subsection{Effect of scaffold density on equilibrium}
\label{subsec:S-agg}
\begin{figure} 
\centering
\subfigure[\null] 
{   \label{fig:S}       \includegraphics[width=7.5cm]{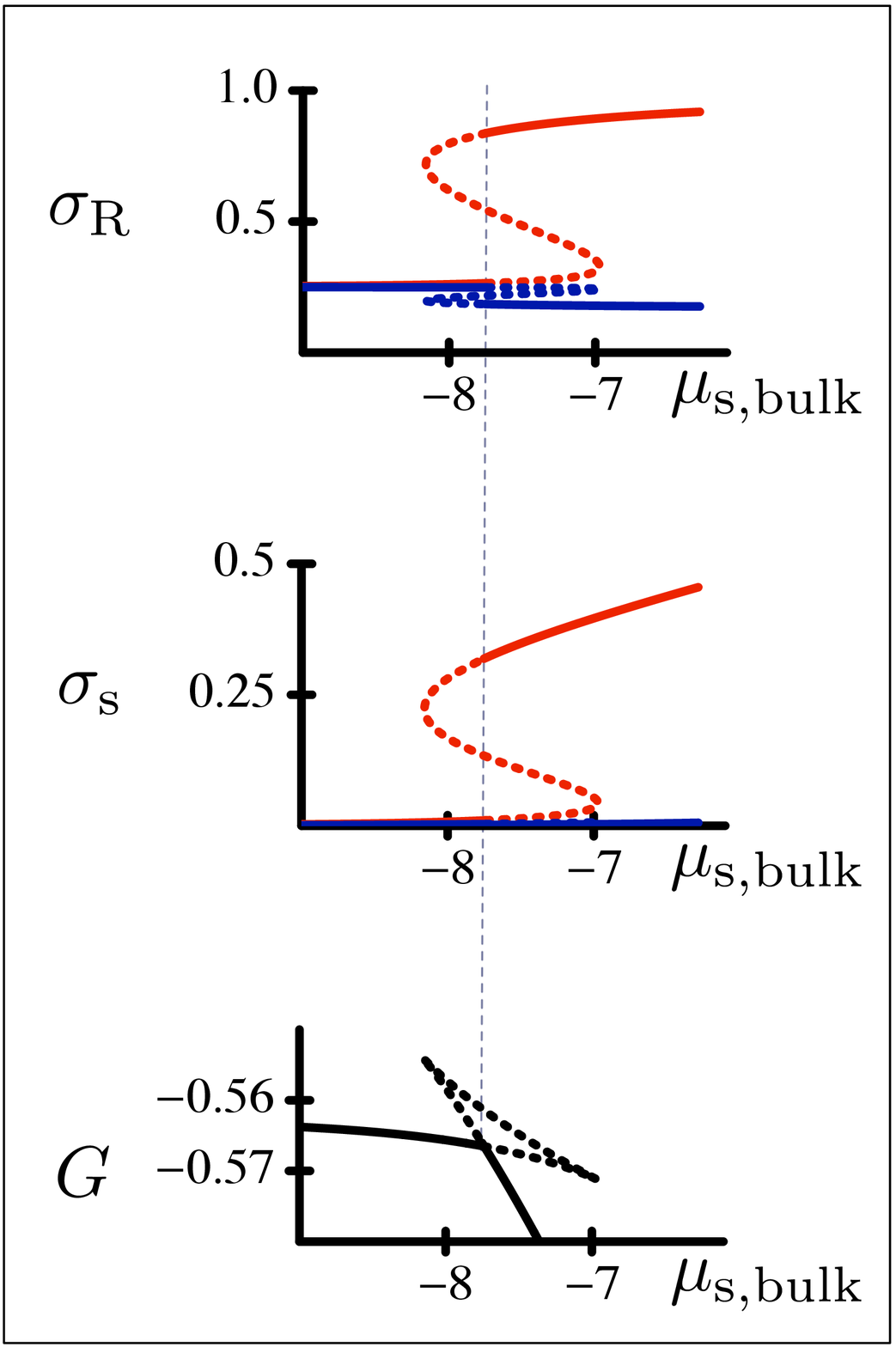}}
\hspace{1cm}
\subfigure[\null] 
{\label{fig:H}     \includegraphics[width=7.5cm]{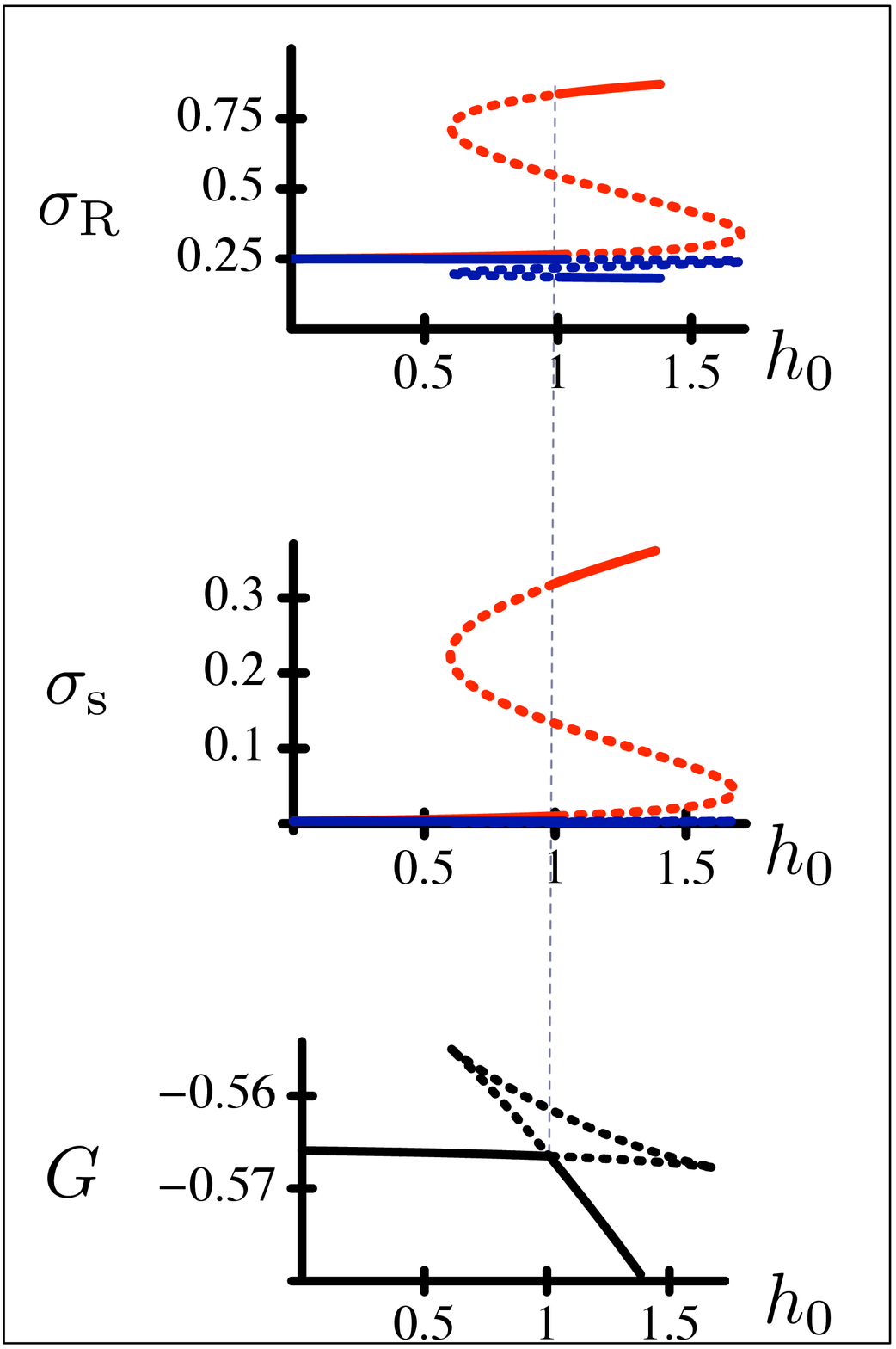}}
\caption{(Color online) 
 (a) Transition (switching) induced by the chemical potential of the scaffold protein in the bulk 
cytoplasm, $\mu_{\rm s,bulk}$ (horizontal axis). (The value of $h_0$ is fixed at $h_0=1.$)
{\it Top ($\sigma_{\rm R}$)}: Densities of the 
membrane receptors in the membrane layer. 
{\it Middle ($\sigma_{\rm s}$)}: Densities of the scaffold proteins in the sub-membrane layer. 
The red [blue] curves represent, respectively, the densities in the synaptic [extrasynaptic] 
zones. 
{\it Bottom ($G$)}: Free energy of the system. The vertical dashed line passing through the figures marks the 
point of phase change, to switch the branch of solutions. Those parts represented by dashed curves are not 
realizable as quasi-equilibrium. \\
(b) Switching induced by the trans-membrane signal, $h_0$. 
(The value of $\mu_{\rm s,bulk}$ 
is fixed at $\mu_{\rm s,bulk}=-7.747$.)
} 
\label{fig:eq-of-state} 
\end{figure}

 We first examine the consequences of the chemical potential of the scaffold proteins in the 
bulk cytoplasm,  $\mu_{\rm s,bulk}$ (Fig.~\ref{fig:eq-of-state}(a)).
As it varies, it modifies the densities of the receptors $\{\sigma^{\rm(z)}_{\rm R},\sigma^{\rm(x)}_{\rm R}\}$ 
 in the respective zones (Fig.~\ref{fig:S}$\sigma_{\rm R}$), 
 and those of the scaffold proteins $\{\sigma^{\rm(z)}_{\rm s},\sigma^{\rm(x)}_{\rm s} \}$
  in the sub-membrane layer (Fig.~\ref{fig:S}$\sigma_{\rm s}$).
The chemical potential $\mu_{\rm s,bulk}$ cannot be defined as an absolute
number, but its  
variation contains the meaning: the higher its value, the more concentrated
the scaffold proteins  
in the bulk layer.

As seen on the curve, there is a region of $\mu_{\rm s,bulk}$ values where
three solutions can  
be found with corresponding values of $G$. Among these, the one corresponding
to the equilibrium  
was determined as that where $G$ has the minimum value for a given $\mu_{\rm
  s,bulk}$, or a given density  of cytoplasmic scaffold protein. 
The selected solutions are shown by solid curves in the figures. 
For completeness, 
 Maxwell's construction is briefly summarized in the rest of this
 subsection.
When following a curve for the density  $\sigma_{\rm R}^{\rm (z)}$ (e.g.
on Fig.~\ref{fig:S}$\sigma_{\rm R}$)  from the minimum value of $\mu_{\rm s,bulk}$ (left-end) to the maximum 
(right-end), there is a portion where  $\mu_{\rm s,bulk}$ decreases.  
This phenomenon occurs simultaneously 
for all the density variables, 
$\sigma_{\rm R}^{\rm (z)}$ and $\sigma_{\rm  R}^{\rm (x)}$ in  
Fig.~\ref{fig:S}$\sigma_{\rm R}$, 
$\sigma_{\rm s}^{\rm (z)}$ and $\sigma_{\rm s}^{\rm (x)}$ in
Fig.~\ref{fig:S}$\sigma_{\rm s}$.  
It applies also to the curve of $G$ (Fig.~\ref{fig:S}$G$). 
The portion of the curve where $\mu_{\rm s,bulk}$ decreases corresponds to the 
branch where the value of $G$ is maximum among the three points corresponding to the  {\it same}
 value of $\mu_{\rm s,bulk}$. The maximum in $G$ implies
the minimum in the probability of realization $\propto e^{-G/\kT}$.
 The portion of 
the curve where the value of $\mu_{\rm s,bulk}$ decreases thus corresponds neither to an equilibrium nor to a 
metastable equilibrium. 
So we exclude this portion of the curves of Fig.~\ref{fig:S}$\sigma_{\rm R}$ 
and  Fig.~\ref{fig:S}$\sigma_{\rm s}$. 

The crossing point 
in Fig.~\ref{fig:S}$G$ indicates the situation where two equilibria can occur with the same probability. 
The solution branches are to be switched at this crossing point. 
The equilibrium densities corresponding to this point can be identified in Fig.~\ref{fig:S}$\sigma_{\rm R}$ and 
Fig.~\ref{fig:S}$\sigma_{\rm s}$.
The switching indicates a discontinuous transition %
 of  mode of the partitioning receptors and scaffold 
proteins between extrasynaptic and synaptic zones. 
This redistribution is a {\it phase change} in the sense that we discussed in
\S~\ref{subsec:chemical-potential}. 
The situation is schematically shown in Fig.~\ref{fig:compartementation}. 
In one phase, which we call the {\it nonlocalized phase}, 
the receptors are found at almost the same density in synaptic and 
extrasynaptic zones, while there is no accumulation of scaffold proteins. 
In the other phase, which we call the  {\it localized phase}, receptors accumulate abundantly in the synaptic zone, 
and are diluted in the extrasynaptic zone. 
And the scaffold proteins also accumulate in the synaptic zone. 
This dramatic contrast in density is genuinely collective 
in the sense that we have {carefully} chosen the parameters of the model so that no phase change takes 
place without reciprocal coupling between the receptors and scaffold proteins, 
$g^{(\alpha)}_{\rm mem-sub}$.   
That is, despite the attractive interaction among the scaffold proteins,
$U_{\rm s}(\sigma_{\rm s})$,  
promoting the accumulation of the scaffold proteins, and the trans-membrane
signal, $(-h_0)$,  
favoring their density in the synaptic zone, 
 they are not enough to realize the distinct accumulation of molecules at the  
synaptic zone if $g^{(\alpha)}_{\rm mem-sub}\equiv 0$.   .  
 In other words, the accumulation would not occur if there were no receptors
 on the membrane.  

\subsection{Effect of trans-membrane signal on equilibrium}
\label{subsec:H-agg}
The trans-membrane signal imposed by the presynaptic element 
{specifies} 
 the organization of the postsynaptic plasma membrane.  
This determines the locus where receptors 
are to accumulate, and is likely to induce an initial metastable state for the formation of the 
synapse. 
In this second study, we therefore analyze the effect of the amount of this trans-membrane signal, $h_0$. 
Fig.~\ref{fig:H} shows the densities of the receptors in the respective zones, similar to Fig.~\ref{fig:S} 
 when changing $h_0$. 
Again, by monitoring the values of $G$, the phase change is identified as the 
self-crossing point of $G$. 
Because of the collective effect, a continuous (quasi-equilibrium) increase of the 
signal $h_0$ induces a sudden accumulation of the molecules in the synaptic zone. 

\subsection{Phase diagram}
\label{subsec:diagramme}
\begin{figure}
\centering
\subfigure[\null] 
{   \label{fig:compartementation} 
    \includegraphics[width=4.5cm]{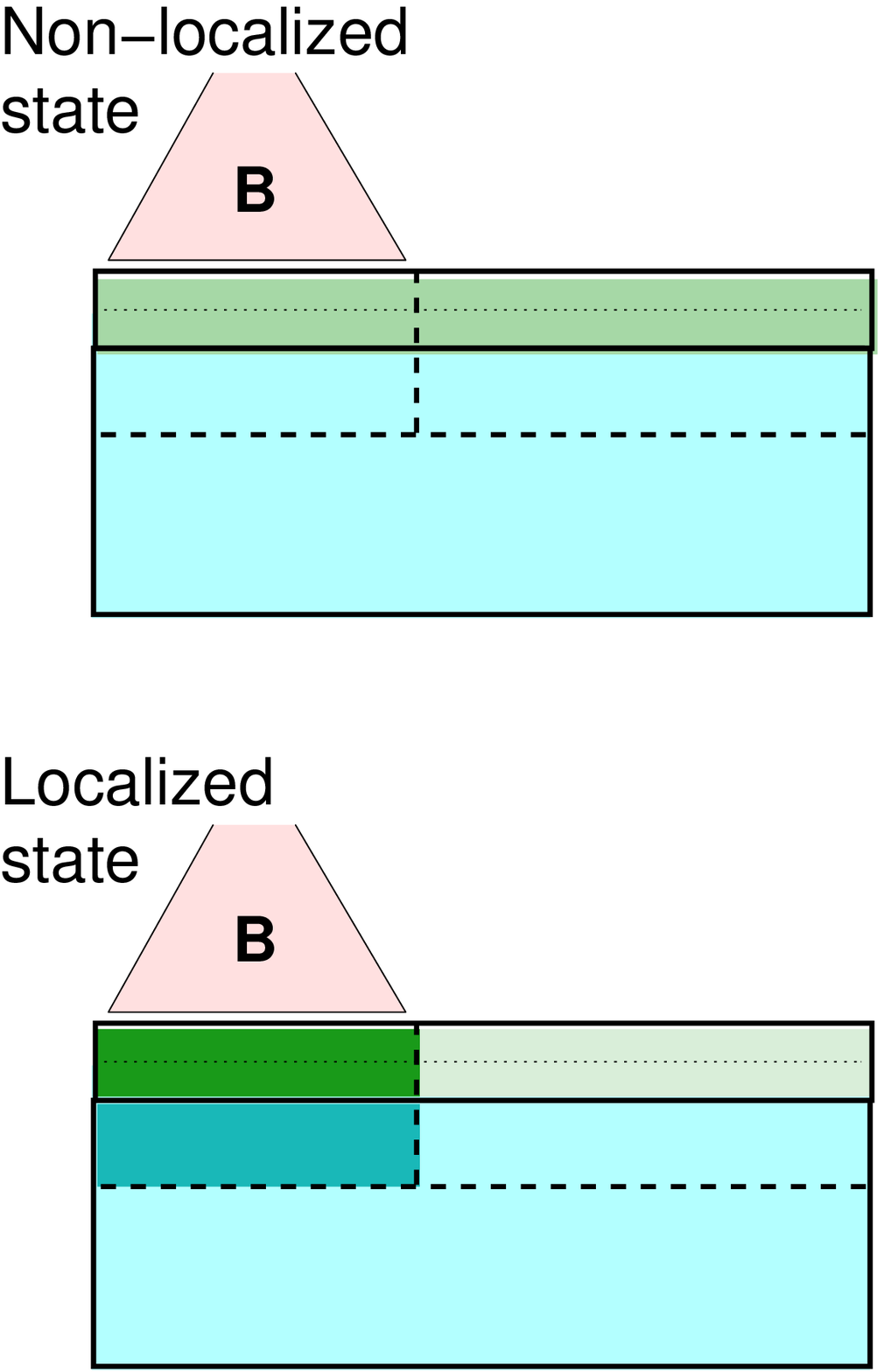}  }\hspace{1cm}
\subfigure[\null] 
{   \label{fig:phase} 
\vspace{-2cm}
    \includegraphics[width=7.5cm] {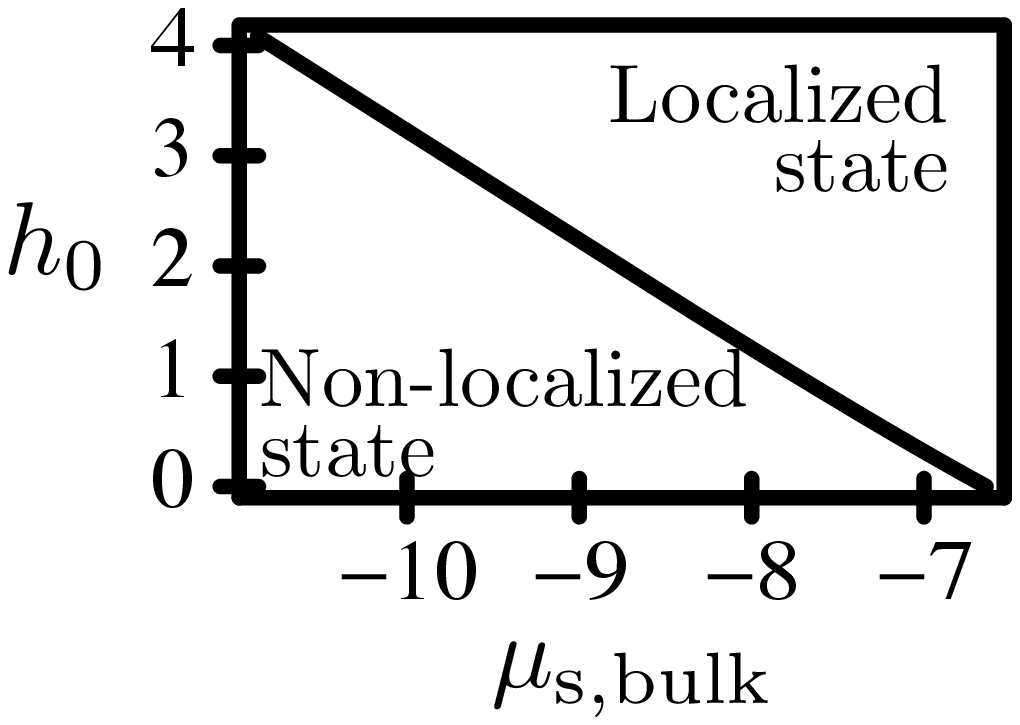}  } 
\caption{(a) (Color online) 
Densities of receptors (in green) and scaffold proteins (in cyan) in the
nonlocalized state (top)  
and localized state (bottom) are shown schematically by concentration of the
colors.  \\
(b) Phase diagram of 
localized {\it vs} nonlocalized phases on the plane of the controlling
parameters.  
The almost straight diagonal curve is the numerical result.
}
\label{fig:combine3} 
\end{figure}

The notion that scaffold 
and adhesion molecules act cooperatively in the formation of the postsynaptic density is 
emphasized in  Fig.~\ref{fig:compartementation}. 
When we allow  both the parameters $\mu_{\rm s,bulk}$ and $h_0$ to vary, 
our main results are summarized in the form of a phase diagram on the
plane of $(\mu_{\rm s,bulk},h_0)$, see  Fig.~\ref{fig:phase}.
This diagram was numerically determined using the technique described in Appendix A.2.   
We observe that the nonlocalized and localized phases are separated by a rather straight
   boundary.
{The reason for this almost straight phase boundary has to be found in the phenomenon of the localization itself. 
Two requirements are to be satisfied. 1) 
in the localized phase
the term  $(\mu_{\rm s,bulk}+h_0)\sigma_{\rm s}^{\rm (z)}$ 
in the free energy $G$ is important while 
$\mu_{\rm s,bulk}\sigma_{\rm s}^{\rm (x)}$ is negligible;
(because $\sigma_{\rm s}^{\rm (x)}\ll \sigma_{\rm s}^{\rm (z)}$), and
2) in the delocalized phase the signal $h_0$ is not important (because $\sigma_{\rm s}^{\rm
   (z)}$ is small).
Therefore,  the sum $(\mu_{\rm s,bulk}+h_0)$ is the term that effectively influences
   the quasi-equilibrium phase.} 

\subsection{{
Non-relevant depletion of extrasynaptic
      receptors upon localization transition}} 
\label{subsec:parametres}
As illustrated in Figs.~\ref{fig:S} and \ref{fig:H}, 
the receptor density at the synapses, $\sigma_{\rm R}^{\rm (z)}$, can 
be localized at the expense of its decrease outside synapse, 
$\sigma_{\rm R}^{\rm (x)}$, when the synapses occupy 
10 percent of the surface, $(A^{\rm (z)},A^{\rm (x)})=(10\%,90\%).$
It is, therefore, of interest to check if the localization transition can
take place if $A^{\rm (z)}$ is much smaller than 
 $A^{\rm (x)}$, e.g.  $(A^{\rm (z)},A^{\rm (x)})=(1\%,99\%),$
in the following two lines of reasonings: 
firstly, the presence of the transition confirms that the decrease in the 
extrasynaptic receptor 
density $\sigma_{\rm R}^{\rm (x)}$ is not  necessary for 
the localization transition, though it may rather be a inhibitory 
factor; 
secondly, the localization transition with a small 
synaptic area, like 1\% of the total membrane, may qualitatively  simulate the initial stage of synaptogenesis.
We have verified numerically that the localization of both the receptors
 and the scaffold proteins occurs even
 with the area fractions, 
$(A^{\rm (z)},A^{\rm (x)})=(1\%,99\%).$ 
The densities $\sigma^{\rm(z)}_{\rm R}$ and $\sigma^{\rm(z)}_{\rm s}$
show a similar jump as in 
Figs.~\ref{fig:S} or \ref{fig:H} while 
$\sigma^{\rm(x)}_{\rm R}$ and $\sigma^{\rm(x)}_{\rm s}$ for the 
extrasynaptic zone display minute change at the localization transition (data
not shown).
With such a small fraction of synaptic area the conservation of 
the total number of receptors, \citer{eq:NtotR}, is 
effectively not a constraining factor, and the persistence of the 
localization transition indicates that the mechanism of the localization 
transition remains in  {\it local} exchanges of molecules
between a synaptic site and its environment.%

\subsection{Effect of weakening of the receptor-scaffold protein interaction}
\label{subsec:vfcrit}%
{
The interaction between the receptors
  and the 
scaffold proteins can be modified by phosphorylation \cite{phosphoryl-Zeta-EMBO07}.
In our model, the  weakening or strengthening of molecular
interactions has effects  on the quasi-equilibrium state of PSD.
It can be simulated  
 by modifying the profile of the 
  function $v(\sigma_{\rm s}^{(\alpha)})=v_{\rm
    f}[1-e^{-v_1\inRbracket{\frac{\sigma_{\rm s}}{\sigma_{{\rm
            s}0}}}-v_2\inRbracket{\frac{\sigma_{\rm s}}{\sigma_{{\rm
            s}0}}}^2}].$ 
 To this aim,
 we varied 
 the global factor, $v_{\rm f}$,  
which accounts for the saturating binding strength. 
We found  (data not shown) (i) that when
 $v_{\rm f}$ is reduced to 70\% of the original value (-6.0 in the units of
our model), the localization transition {\it vs} $\mu_{\rm s,bulk}$ %
almost disappears, while 
the receptor density in the synapse, $\sigma^{\rm(z)}_{\rm
  s},$ has strong non-linear behavior; 
   (ii) furthermore, 
when $v_{\rm f}$ is reduced to 50\% of the original value, there is no more 
localization
transition and   $\sigma^{\rm(z)}_{\rm s}$ displays %
a smooth sigmoidal dependence
on  $\mu_{\rm s,bulk}$.}

\subsection{Limit of robust characters}\label{subsec:compet}
{
The stability of 
 receptor density in the synaptic region $\sigma^{\rm(z)}_{\rm R}$
is an indication of the robustness of the localized state.
This robustness, however, has a limit. 
The quasi-equilibrium state for different (conserved) values of 
the total receptor number, $N_{\rm R}$  (between 0.02 and 0.4 in the arbitrary unit) 
was estimated with 
fixed values of $h_0$ and $\mu_{\rm s,bulk}$.
In the localized state the receptor density in the synaptic region, $\sigma^{\rm(z)}_{\rm R}$ (as well as 
$\sigma^{\rm(z)}_{\rm s}$), is almost saturated and constant while that in the extrasynaptic region increases roughly proportionally to $N_{\rm R}$. 
But if $N_{\rm R}$  is less than a critical value, $N_{\rm R}^{(loc)}\simeq 0.16,$
then the localized state is destroyed and the receptor densities in synaptic and extrasynaptic regions
are almost the same and proportional to  $N_{\rm R}$. 
Therefore, the robustness is closely related to the cooperative effect.
}
{
That the localization disappears for too small value of $v_{\rm f}$
(\S~\ref{subsec:vfcrit}) implies that the robustness is also closely related
to the reciprocal 
stabilization of the PSD.  
}

\section{Discussion} 
\label{sec:discussion}

\subsection{Summary of the results  and comparison with other theories}

In this paper we present a minimal three-layer two-compartment model to describe the formation of the postsynaptic assembly of membrane receptors and scaffold proteins. 
We found the discontinuous phase change between the nonlocalized and localized
phases.  
 In the localized phase, the stable high density of receptors at synaptic sites
is compatible with the mobility of individual receptors. 
 This accounts for the observation that 
synapse formation is almost an all-or-none process, operating on a short
time scale in the range of  the diffusion constant of individual molecules.  
{
(Here one should take into account not only the diffusion of receptors but also the local turnover of scaffold proteins. ) 
We note that  the {\it latency time} for synapse formation should be distinguished from the {\it duration of synapse formation}, which we discuss here.  The former time results from the metastability of the receptor-scaffold assembly. This is indeed one of the main message of this paper (see V.B {\it b} and {\it c} below).
}  
Although our model assumes the quasistatic equilibrium, such decoupling between
kinetics and 
thermodynamics (\S~\ref{subsec:2A}) should also be true even if the system 
is slightly out of equilibrium. 
Such flexibility is the basis of the responsiveness of the synaptic junction
(see, for example, a review   \cite{Misteli}). 
{
Understanding how the number of receptors is determined at steady state as a 
set-point of dynamic 
equilibrium provides the mechanism by which this number can be modified during
plastic changes  of synaptic strength (the gain of information transfer). 
}

{
Recently, a new model has been proposed \cite{shouval} in which the stability of receptor density is 
compatible with individual receptor turnover. This model 
 deals only with the membrane receptor 
zone in the synaptic compartment as we defined it. 
Nevertheless, it accounts for the key idea of 
cooperativity in maintaining the stable density of receptors, as too does our model. 
{However it does not take into account the interaction of receptors with scaffolding molecules nor the chemical potentials resulting from concentration differences in the cellular compartments.} 
Therefore, the 
model we propose complements the concept of cooperativity within a more realistic framework 
based on 
experimental knowledge demonstrating the exchanges between extrasynaptic 
and synaptic receptors \cite{AT-DC-trends}. 
This concept of cooperativity has been suggested to operate between the 
acetylcholine receptor and the 43kD/rapsyn protein \cite{Changeux}. }
{
Recently, Fusi {\it et al.} proposed a 
cascade mechanism to generate different time scales of synaptically stored memories \cite{switch},
which sheds light on the quasi-equilibrium approach that we propose. 
As the kinetics are independent of the stability of postsynaptic molecular construction, different time scales can 
coexist to account for the dynamic turnover of constituent molecules in the postsynaptic density. 
The layered structure of the postsynaptic multi molecular assembly reflects a cascade 
of interactions (the trans synaptic molecule signaling to the scaffold protein
assembly and then receptor accumulation via reciprocal stabilization with the
scaffold proteins).}

\subsection{Implications of the results  and qualitative comparison with experiments}

\paragraph{Collective stabilization justifies the non-stoichiometry.}
{
Only 20 to 30\% of PSD-95, a scaffold protein 
present at excitatory synapses, 
in the sub-membrane layer 
is likely to be bound to receptors at steady 
state \cite{Signal-Kennedy,PSdensity-Sheng}. 
This proportion, well below 100\%, is accounted 
for by our model.
Since the ratios $\sigma^{\rm (z)}_{\rm R} /\sigma^{\rm (z)}_{\rm s}$ and
  $\sigma^{\rm (x)}_{\rm R} /\sigma^{\rm (x)}_{\rm s}$ 
are determined by the reciprocal and collective stabilization, there is no
reason for them to be a rational number. 
 From the values of the densities of receptors 
$(\sigma^{\rm (z)}_{\rm R} , \sigma^{\rm (x)}_{\rm R} )$  and of scaffold proteins 
  $(\sigma^{\rm (z)}_{\rm s} ,\sigma^{\rm (x)}_{\rm s} )$     (in Fig.\ref{fig:eq-of-state}(a) and (b)), 
  we can read out the proportion of receptors interacting with scaffold
  proteins,  
  i.e.  $\sigma^{\rm (z)}_{\rm R} /\sigma^{\rm (z)}_{\rm s}$ or 
  $\sigma^{\rm (x)}_{\rm R} /\sigma^{\rm (x)}_{\rm s}$   
in units of  $\sigma_{{\rm R}0} /\sigma_{{\rm s}0}$ (data not shown). 
  In the synaptic zone, the ratio $\sigma^{\rm (z)}_{\rm R} /\sigma^{\rm
    (z)}_{\rm s}$   increases dramatically 
upon the localization transition, while in the extrasynaptic zone the ratio 
   $\sigma^{\rm (x)}_{\rm R} /\sigma^{\rm (x)}_{\rm s}$  decreases only 
slightly upon the localization. 
  This is due to differences 
in surface area \cite{Signal-Kennedy,PSdensity-Sheng}. }

\paragraph{Competitive binding can destroy the localized phase.}
{
Disturbing molecules (such as ones producing {dominant-negative 
    competitive binding}) modifies the energy profiles 
by altering 
the chemical potential  $\mu_{\rm s,bulk}$. 
 In Appendix.B the equilibrium theory of competitive binding is summarized briefly.
The theory shows that the competitive molecule species (e.g. B) 
versus the principal species (e.g. A) 
  effectively reduces the chemical
potential of the latter, $\mu_{\rm A}^0$ by a quantity 
$\Delta \mu_{\rm A}^0=-\kT \ln[1+e^{(U_{\rm B}+\mu_{\rm B}^0)/\kT}],$ where
$U_{\rm B}$ and $\mu_{\rm B}^0$ are
the binding energy and
the external chemical potential, respectively, 
for the competitive/dominant-negative molecule. 
As we found that the low chemical potential $\mu_{\rm s,bulk}$ destabilizes
the localized phase, we predict  that the competitive binding with scaffold
proteins tends to destabilize the localized phase. 
}

\paragraph{The fate of PSD after sudden disappearance of localization signal
  should depend non-linearly on the cytoplasmic scaffold protein concentration.}
{
Although our approach is quasistatic, we 
can draw some conclusions 
about the non-quasistatic phenomena since the  
response of the postsynaptic density (PSD) to a sudden 
disappearance  of the localization signal, $h_0$, should depend
 on the other parameters of the system 
(see \cite{h0devel-BrRes81} for synapses during development and
\cite{sotelo-brain-res73,denervation-AT-JN92} for mature synapses).
As seen on the phase diagram, Fig.7:(b), 
 the localization transition occurs
even when $h_0=0$ if the concentration of the scaffold protein is
 large enough (or, $\mu_{\rm s,bulk}\ge -6.3$ in Fig.7:(b)).
For $\mu_{\rm s,bulk}$ near this threshold value, the sudden disappearance  of $h_0$ will leave, at least transiently, the PSD as a (meta)stable state for $h_0=0$.
However, if $\mu_{\rm s,bulk}$ was far below the threshold value,  then
the aggregate will be disrupted rapidly by lateral diffusion %
after the disappearance  of $h_0$.
In conclusion we predict that
 the life-time of the PSD after the sudden disappearance of $h_0$ 
depends on $\mu_{\rm s,bulk}$ in a highly non-linear manner. 
The detailed dynamic response, however, 
 is beyond the scope of the present quasi-equilibrium framework of our
paper.
}

\paragraph{Delayed time for the construction of a new synapse can be due to
  the metastable nonlocalized phase.}
{
A complementary issue to the above paragraph is 
{``how long would a new synapse take to  assemble?''}
Experimentally, the assembly of a new PSD 
takes at least tens of minutes, 
more likely 1-2 hours \cite{Friedman-neuron00},
{which is not rapid}, given the characteristic 
diffusion constant of individual receptors (in the order 
of $10^{-2}\mu$m$^2$/sec). 
This time lag supports our model of cooperative 
interaction underlying synaptic localization of receptors.
 When the expression of the scaffold proteins in the cytoplasm
raises $\mu_{\rm s,bulk}$ just up to the localization transition point,
the nonlocalized state remains still metastable.
Under such conditions the clustering of PSD must wait for  
the random rare event (``nucleation'') which assembles a critical
concentration of  
receptors as well as scaffold proteins. 
We then predict that the waiting time of the nucleation 
should be stochastically distributed, typically obeying an exponential
distribution.  }

\paragraph{The model accounts for the triggering role of trans-membrane signal
  on the localization.} 
{
The phenomenon of localization 
 could be intuitively {postulated} from the known molecular interactions, 
for example, between neuroligin and the scaffold protein PSD-95
\cite{Scheiffele03}.   
Experimental data indicate that the neurexin-neuroligin heterophilic  
interaction induces the formation of the postsynaptic micro-domain
\cite{Scheiffele03}, and that, once it begins, 
{it is a rapid phenomenon}, taking place within minutes \cite{Basarsky94}.
The present model is consistent with these observations. 
That is, the formation of postsynaptic micro-domains is 
almost an all-or-none phenomenon involving a phase change, and is imposed by
the presynaptic contact.  }

\paragraph{The model admits the spontaneous formation of sub-membrane aggregates.}
{
In the early period of synaptogenesis spontaneous formation of sub-membrane aggregates of scaffold proteins have been observed, notably at the locations of dendrite-dendrite contact or dendrite-substrate contact\cite{Colin-JCompara96,Colin-JCompara98}.
In our model, spontaneous localization of scaffold
proteins can be realized without receptors or without the transsynaptic bias,
$h_0$,  if we modify the parameter characterizing the attractive interaction
among scaffold proteins, that is, $|U_2|$ in $U_{\rm s}(\sigma_{\rm s})$ (see
(\ref{fig:Us})). }

\subsection{Future problems}
\label{subsection:adapt}

{
{As future problems we should incorporate other factors that might exert
influence on synaptic receptor clustering. In particular, we may take into account
the mechanism involving aggregation of receptors through direct interaction with an extracellular-matrix molecule \cite{extracell-Dityatev-Nature03},
the activation of receptors which is indirectly related to the electrodiffusion of charged neurotransmitter molecules \cite{Efields-Sylantyev-Science08},
and the dendritic spine geometry (volume of spine head and spine length),
which is strongly correlated with the number of receptors on the spine
\cite{spine-geometry-NatNeur01}.}
}

{
An important question is how much time an individual receptor spends in the synaptic zone. 
At steady state, the fraction of time spent by a particular receptor on a particular 
synaptic contact should be proportional to the density of the receptors at the contact. 
This is true if all receptors are well mixed so that there is no separation between the permanently immobile receptors and mobile receptors. 
Experimentally, single-particle tracking measurements have established that about half of the
receptors are mobile at central excitatory synapses  \cite{DC-AT}.  
In contrast, FRAP  experiments of glutamate receptors at Drosophila neuromuscular junctions suggest that they are immobilized once they enter into 
the postsynaptic domain \cite{Glu-Dyn-Rasse}. 
Models to assess these observations must go beyond the simple dichotomy of
synaptic - vs extrasynaptic - zones.
}

{
A major unsolved problem is the determining mechanism of the postsynaptic
micro-domain. 
The size of this domain, 
although variable, is maintained in a relatively narrow range, 100-300  nm in diameter \cite{Peters76}. 
In double transfection experiments with glycine receptor and its associated 
scaffold proteins, it was found that the aggregates
of scaffold proteins had a size 
close to that of postsynaptic
micro-domains \cite{Meier-JCellS00} even in the absence of presynaptic terminals.
However, 
this will not specify the size of the localized cluster of scaffold proteins.
}
One may conjecture several different mechanisms for the regulation of the size of postsynaptic micro-domains. 
A cost of curvature driven energy of a microdomain structure  might define 
an optimal size of aggregates as found for clathrin-coated
vesicle formation\cite{clathrin_JCS97}.
Or, the steric repulsion among molecules reflecting their three-dimensional arrangement may limit the size of the cluster \cite{Size-Lang-Sci07}.

\section*{Acknowledgments}
We acknowledge Jean-Fran\c{c}ois Joanny for reading the original manuscript
and providing a few references.
We also acknowledge a referee for bringing  numerous recent references
to our attention.
This work was supported by a grant from the Agence Nationale de la  Recherche,
ANR 05-Neur-043-02.

\mbox{}\\
\section*{Appendix A. Technical notes}
\subsection*{A.1 Brief summary of the Lagrange multiplier method}
\label{subsec:Lagrange}
This method finds stationary points (local maxima etc.) of $f({\bm{x}})$ with
the 
constraint $g({\bm{x}})=0$, where 
${\bm x}=(x_1,\ldots,x_n)\equiv \{x_i\}$. 
A point of stationary point, ${\bm x}^*$, 
together with a constant called the Lagrange multiplier, $\lambda$,
 must satisfy the following condition:
\beq
g({\bm{x}^*})=0,\quad 
\left.\frac{\partial (f-\lambda g)}{\partial x_i}
\right|_{\bm{x}=\bm{x}^*}=0,
\quad \mbox{$i=1,\ldots,n$}.
\eeq
The reason is that at ${\bm x}^*$ the contour
surface of $f(\bm{x})=f(\bm{x}^*)$ and that of $g(\bm{x})=0$
must share the same tangential plane, and that,
for any function, say $\phi(\bm{x})$, the normal vector 
of a tangential plane is 
along $(\partial \phi/\partial x_1,\ldots,\partial \phi/\partial x_n)$,
which can be easily verified in the case of a line $ax_1+bx_2=c$.

\subsection*{A.2 Numerical solution procedure}
\label{subsec:app-solve}
Formally, the problem is to solve $n$  coupled non-linear equations
for $(n+1)$ variables, 
$f_i(x_1,\ldots,x_n,x_{n+1})=0$ ($i=1,\ldots,n$).
Once we have {\it a} particular solution $(x_1,\ldots, x_n,x_{n+1})$, then
 we may use the differential equations describing 
the solution curve in the space of $\bm{x}\equiv(x_1,\ldots, x_n,x_{n+1})$:
$\sum_{j=1}^{n+1} {\sf M}_{ij}\rmd\bm{x}_j =0$  ($i=1,\ldots,n$), 
where ${\sf M}$ is the $n\times(n+1)$ matrix containing the components, 
${\sf M}_{ij}\equiv {\partial f_i}/{\partial x_j}$  ($i=1,\ldots, n$ and $j=1,\ldots,n+1$).
The latter equations can be solved using the cofactor of ${\sf M}$, 
which we denote by $\tilde{\sf M}$ (i.e., $\tilde{M}_{i,j}$ is $(-1)^{i+j}$ 
times the minor entry of $M_{i,j}$): 
\beq
\frac{\rmd \bm{x}}{\rmd s}=
(\tilde{\sf M}_{n+1,1},\ldots,\tilde{\sf M}_{n+1,n+1})^t,
\eeq
where $s$ is a parameter along the solution curve.

In the context of solving \citer{eq:NtotR} and  \citer{eq:cond-eq}
we have $n=5$.
The variable is
$\bm{x}=(\sigma^{\rm(z)}_{\rm R},\sigma^{\rm(x)}_{\rm R},\sigma^{\rm(z)}_{\rm
  s}, \sigma^{\rm(x)}_{\rm s}, \mu_{\rm R}^*,\xi),$
where the sixth component $\xi$ stands for either the parameter 
$\mu_{\rm s,bulk}$ (\S~\ref{subsec:S-agg}) or $h_0$ (\S~\ref{subsec:H-agg}).
To find the phase boundary (\S~\ref{sec:discussion}, 
Fig.~\ref{fig:phase}), we have
$n=11$, i.e. twice the five conditions of 
\citer{eq:NtotR} and  \citer{eq:cond-eq} for
each phase, 
plus the equality of the total free energy, $G$.
The variables $\bm{x}$ consists of twice the five variables,
$\{\sigma^{\rm(z)}_{\rm R},\sigma^{\rm(x)}_{\rm R},\sigma^{\rm(z)}_{\rm s},
\sigma^{\rm(x)}_{\rm s}, \mu_{\rm R}^* \}$, for the coexisting phases, plus
$\mu_{\rm s,bulk}$ and $h_0$.

\section*{Appendix B: Effect of competitive
  binding  }
\label{app-antago}
We take as the Helmholtz free energy 
$F/\kT= -n_{\rm A}\frac{U_{\rm A}}{\kT}-n_{\rm B}\frac{U_{\rm B}}{\kT}+$
$n_{\rm A}\ln (n_{\rm A}/n)+$ $n_{\rm B}\ln (n_{\rm B}/n)+$ 
$n_{\rm V}\ln(n_{\rm V}/n),$ where $n_{\rm A}$ and 
$n_{\rm B}$ are the number of the A [B] molecules occupying among 
the $n$ binding sites, respectively, and $n_{\rm V}=n-n_{\rm A}-n_{\rm B}.$
We impose the chemical equilibrium conditions with the solvent chemical potentials
for A and B, which we denote by $\mu_{\rm A}^0$ and $\mu_{\rm B}^0,$
respectively;
$\mu_{\rm A}^0=\partial F/\partial n_{\rm A}$ and $\mu_{\rm B}^0=\partial
F/\partial n_{\rm B}$. 
In the absence of B molecules (i.e. $\mu_{\rm B}^0=-\infty$), equilibrium
condition for the A molecule 
binding writes $\mu_{\rm A}^0=-U_{\rm A}+\kT\ln[n_{\rm A}/(n-n_{\rm A})]$,
while for finite $\mu_{\rm B}^0$, the right hand side of this condition is
shifted by $-\Delta \mu_{\rm A}^0$ ($>0$), where
$\Delta \mu_{\rm A}^0\equiv -\kT \ln[1+e^{(U_{\rm B}+\mu_{\rm B}^0)/\kT}].$
This implies that the attractive energy $-U_{\rm A}$ for A molecule is  
partly cancelled by this amount due to the competitive/dominant-negative
molecules, B.  
If $U_{\rm B}\ll \kT$, the effect is small, in the order $\kT$ (more precisely
$\simeq -\kT e^{(U_{\rm B}+\mu_{\rm B}^0)/\kT}$). 
Contrastingly, large $U_{\rm B}/\kT$ has 
a strong influence of the competing molecules due to 
 the interference,
$\Delta \mu_{\rm A}^0\simeq - (U_{\rm B}+\mu_{\rm B}^0)$.

\section*{Appendix C. Note for the biologists}
In this appendix we explain in general terms, easily understandable for
biologists, the object of the modelling accounting for the compatibility
between synaptic stability and molecular mobility.

{ The stability of the synaptic structure, with its mobile receptors, is a
 complex matter, because the local turnover (at synapses) of the constituent
 elements is shorter than the lifetime of the synapse  (see comment by
 \cite{ShengParadox}). In the light of the dynamics of individual molecules
 such as diffusion in the plane of the plasma membrane for receptors and of
 spatial 3D diffusion of scaffolding molecules in the cytosol, it was
 necessary to establish a theoretical background accounting for the
 accumulation of receptors at synapses. The present model has been developed
 including the extrasynaptic membrane.} 

{It  stresses the {\it quasi-equilibrium} which is valid on a time scale
  shorter  
than that of receptor turnover on the membrane. 
It is not known if the turnover by exocytosis and 
endocytosis promotes exchange of receptors between the synaptic and extrasynaptic 
zones, or whether such active exchange has a role on large time scales.
However, this raises 
the question of multi molecular assembly as a global entity in which regulation can operate without 
destroying the integrity of the structure. 
In more biological terms, the important question is 
how molecules such as receptors or scaffold proteins can be added or removed while maintaining the 
synaptic function with variable gain. 
The present model provides a general framework in which it 
is now possible to conceive of molecular interactions in terms of chemical
potentials and, therefore, 
to model a kinetic 
 view of the synaptic multi-molecular assembly. 
It is also expected that the model we propose will allow a unification of the different levels of 
postsynaptic events, from the chemical interaction between receptors and
scaffolding molecules up  
to the plasticity of synaptic transmission. 
In this context we mention three aspects which may help refine  
  our  study in the future: {heterogeneity of time scales},
  {collective stabilization, adaptation and molecular exploration upon 
PSD formation}.

{The components used for the modelling are of the same nature 
as those used  in physical chemistry to account for the thermodynamics of
chemical reactions, which also holds 
 in living system.
 }
{The model predicts a discontinuous increase of the density of receptors at the  synaptic contact through the transition to the localization regime. 
Unless there is an 
unusual kinetic mechanism to increase the mobility of individual receptors during the localization 
transition, the increase of receptor density in a synaptic zone should also imply a lengthening of the 
residence time of individual receptors. 
However, one should stress that the stabilization of receptor density (number
of receptors) in the synaptic zone with indefinite lifetime is compatible 
with a finite residence time of an individual receptor on a synaptic site. }
Thus,
{ the persistence of the individual  
mobility of receptors facilitates fast adaptation of receptor numbers in
relation to changes in  neuronal activity. 
}

{Another concept which arises 
from the present model is the notion that stabilization is a {\it reciprocal mechanism}. 
 In other terms, scaffold proteins stabilize receptors, and receptors stabilize 
scaffold proteins. 
This means that the local turnover of a given protein is not likely by itself to determine 
the turnover of the structure. 
In the context of synaptogenesis, reciprocity ensures the synchronized and
adaptive 
construction of the synapses, since neither receptor nor scaffold protein
nor transsynaptic interaction alone can stabilize the localization. 
Reciprocity introduces robustness against the fluctuations in total receptor
number associated with exo-/endocytosis at extrasynaptic sites. 
In addition, the reciprocity is likely to attenuate the amplitude 
of stochastic fluctuations of the receptor numbers at each synaptic site.

{ Another major outcome of the proposed model is that it accounts for changes during synaptic 
plasticity or even during synapse formation, which may result from changes in receptor number in the plasma membrane and/or from changes in the density of scaffold proteins in the 
cytosol. It explains how changes in densities, i.e. chemical potentials, of
receptors and scaffold proteins 
lead to a new steady state of the postsynaptic molecular assembly: the {\it cooperativity} underlying 
the discontinuous change in density distributions allows the system to switch from one point of 
equilibrium (set-point) to another one, by small changes in key parameters
(trans-synaptic signal, cytoplasmic  
density of scaffold proteins, density of extrasynaptic receptor). 
At the molecular level, the mechanisms for the stoichiometry of interaction of
receptors with individual scaffold proteins are not fully understood.
}

{
 The model is consistent with the fact
that, once the formation of synaptic contacts starts,
it is likely to be a rapid process as the system is cooperative and almost
auto-catalytic.}  
{
That the 
recruitment kinetics of various PSD  molecules are 
remarkably similar indicates that PSD assembly rate is governed by a common
upstream rate-limiting process  \cite{Bresler04}. 
In this context it has been observed that the receptor and scaffold
proteins can be already associated on the extrasynaptic membrane
\cite{Erensperger-BJ07}. 
Intracellular packages of NMDA receptors (NMDA-R) or AMPA receptors
(AMPA-R) with the scaffold protein PSD-95 have been identified
\cite{El-Husseini1-JCB00,El-Husseini1-Cell02}.
Also  packages of glycine receptor (Gly-R) and its
scaffold protein partner, gephyrin, were found to be 
transported through the secretion pathway from the Golgi apparatus 
to the membrane \cite{Cyril-JNS04}.
Therefore two mechanisms are cooperative for the assembly of a new PSD:
firstly, as mentioned above, pre-assembly of receptor-scaffold complexes in the
secretion pathway \cite{cyrilAT-JN04}, 
secondly, the high diffusion rate of the receptors, which makes
them explore large areas of plasma membrane 
(\cite{ATCDrev08} and the references cited therein). Therefore,
molecules at any location of the
cell surface may encounter with a high frequency. }
As a consequence, a local trans-synaptic interaction 
creates a potential well that
will rapidly trap the diffusing molecules.
}
{These chemical kinetics have to be reconciled with specific biological 
mechanisms. This can now be achieved because the behavior of individual
 molecules can be monitored (see \cite{ATCDrev08}), 
therefore allowing access to mechanisms normally
 hidden in the convoluted statistics  
of the behavior of large numbers of molecules. }

\bibliographystyle{apsrev.bst}    
\bibliography{KS_AT_refs}

\begin{thebibliography}{66}
\expandafter\ifx\csname natexlab\endcsname\relax\def\natexlab#1{#1}\fi
\expandafter\ifx\csname bibnamefont\endcsname\relax
  \def\bibnamefont#1{#1}\fi
\expandafter\ifx\csname bibfnamefont\endcsname\relax
  \def\bibfnamefont#1{#1}\fi
\expandafter\ifx\csname citenamefont\endcsname\relax
  \def\citenamefont#1{#1}\fi
\expandafter\ifx\csname url\endcsname\relax
  \def\url#1{\texttt{#1}}\fi
\expandafter\ifx\csname urlprefix\endcsname\relax\def\urlprefix{URL }\fi
\providecommand{\bibinfo}[2]{#2}
\providecommand{\eprint}[2][]{\url{#2}}

\bibitem[{\citenamefont{Malinow and Malenka}(2002)}]{Malinov02}
\bibinfo{author}{\bibfnamefont{R.}~\bibnamefont{Malinow}} \bibnamefont{and}
  \bibinfo{author}{\bibfnamefont{R.~C.} \bibnamefont{Malenka}},
  \bibinfo{journal}{Annu. Rev. Neurosci.} \textbf{\bibinfo{volume}{25}},
  \bibinfo{pages}{103} (\bibinfo{year}{2002}).

\bibitem[{\citenamefont{Sheng and Kim}(2002)}]{ShengKim}
\bibinfo{author}{\bibfnamefont{M.}~\bibnamefont{Sheng}} \bibnamefont{and}
  \bibinfo{author}{\bibfnamefont{M.~J.} \bibnamefont{Kim}},
  \bibinfo{journal}{Science} \textbf{\bibinfo{volume}{298}},
  \bibinfo{pages}{776} (\bibinfo{year}{2002}).

\bibitem[{\citenamefont{Bredt and Nicoll}(2003)}]{Bredt03}
\bibinfo{author}{\bibfnamefont{D.~S.} \bibnamefont{Bredt}} \bibnamefont{and}
  \bibinfo{author}{\bibfnamefont{R.~A.} \bibnamefont{Nicoll}},
  \bibinfo{journal}{Neuron} \textbf{\bibinfo{volume}{40}}, \bibinfo{pages}{361}
  (\bibinfo{year}{2003}).

\bibitem[{\citenamefont{Verkhovsky et~al.}(1999)\citenamefont{Verkhovsky,
  Svitkina, and Borisy}}]{verkhovsky}
\bibinfo{author}{\bibfnamefont{A.}~\bibnamefont{Verkhovsky}},
  \bibinfo{author}{\bibfnamefont{T.}~\bibnamefont{Svitkina}}, \bibnamefont{and}
  \bibinfo{author}{\bibfnamefont{G.}~\bibnamefont{Borisy}},
  \bibinfo{journal}{Current Biol.} \textbf{\bibinfo{volume}{9}},
  \bibinfo{pages}{11} (\bibinfo{year}{1999}).

\bibitem[{\citenamefont{Alberts et~al.}(2002)\citenamefont{Alberts, Johnson,
  Lewis, Raff, Roberts, and Walter}}]{rev-kinesin}
\bibinfo{author}{\bibfnamefont{B.}~\bibnamefont{Alberts}},
  \bibinfo{author}{\bibfnamefont{A.}~\bibnamefont{Johnson}},
  \bibinfo{author}{\bibfnamefont{J.}~\bibnamefont{Lewis}},
  \bibinfo{author}{\bibfnamefont{M.}~\bibnamefont{Raff}},
  \bibinfo{author}{\bibfnamefont{K.}~\bibnamefont{Roberts}}, \bibnamefont{and}
  \bibinfo{author}{\bibfnamefont{P.}~\bibnamefont{Walter}},
  \emph{\bibinfo{title}{Molecular Biology of the Cell}}
  (\bibinfo{publisher}{Garland Science (New York)}, \bibinfo{year}{2002}),
  \bibinfo{edition}{4th} ed.

\bibitem[{\citenamefont{Sekimoto et~al.}(2004)\citenamefont{Sekimoto, Prost,
  J{\"{u}}licher, Boukellal, and Bernheim-Grosswasser}}]{brisure}
\bibinfo{author}{\bibfnamefont{K.}~\bibnamefont{Sekimoto}},
  \bibinfo{author}{\bibfnamefont{J.}~\bibnamefont{Prost}},
  \bibinfo{author}{\bibfnamefont{F.}~\bibnamefont{J{\"{u}}licher}},
  \bibinfo{author}{\bibfnamefont{H.}~\bibnamefont{Boukellal}},
  \bibnamefont{and}
  \bibinfo{author}{\bibfnamefont{A.}~\bibnamefont{Bernheim-Grosswasser}},
  \bibinfo{journal}{Eur. Phys. J. E} \textbf{\bibinfo{volume}{13}},
  \bibinfo{pages}{247} (\bibinfo{year}{2004}).

\bibitem[{\citenamefont{Kruse et~al.}(2004)\citenamefont{Kruse, Joanny,
  Julicher, Prost, and Sekimoto}}]{kerato5}
\bibinfo{author}{\bibfnamefont{K.}~\bibnamefont{Kruse}},
  \bibinfo{author}{\bibfnamefont{J.-F.} \bibnamefont{Joanny}},
  \bibinfo{author}{\bibfnamefont{F.}~\bibnamefont{Julicher}},
  \bibinfo{author}{\bibfnamefont{J.}~\bibnamefont{Prost}}, \bibnamefont{and}
  \bibinfo{author}{\bibfnamefont{K.}~\bibnamefont{Sekimoto}},
  \bibinfo{journal}{Phys. Rev. Lett.} \textbf{\bibinfo{volume}{92}},
  \bibinfo{pages}{078101} (\bibinfo{year}{2004}).

\bibitem[{\citenamefont{Kruse et~al.}(2005)\citenamefont{Kruse, Joanny,
  Julicher, Prost, and Sekimoto}}]{kerato5bis}
\bibinfo{author}{\bibfnamefont{K.}~\bibnamefont{Kruse}},
  \bibinfo{author}{\bibfnamefont{J.-F.} \bibnamefont{Joanny}},
  \bibinfo{author}{\bibfnamefont{F.}~\bibnamefont{Julicher}},
  \bibinfo{author}{\bibfnamefont{J.}~\bibnamefont{Prost}}, \bibnamefont{and}
  \bibinfo{author}{\bibfnamefont{K.}~\bibnamefont{Sekimoto}},
  \bibinfo{journal}{Eur. Phys. J. E} \textbf{\bibinfo{volume}{16}},
  \bibinfo{pages}{5} (\bibinfo{year}{2005}).

\bibitem[{\citenamefont{Triller and Choquet}(2005)}]{AT-DC-trends}
\bibinfo{author}{\bibfnamefont{A.}~\bibnamefont{Triller}} \bibnamefont{and}
  \bibinfo{author}{\bibfnamefont{D.}~\bibnamefont{Choquet}},
  \bibinfo{journal}{Trends Neurosci.} \textbf{\bibinfo{volume}{28}},
  \bibinfo{pages}{133} (\bibinfo{year}{2005}).

\bibitem[{\citenamefont{Moss and Smart}(2001)}]{Moss01}
\bibinfo{author}{\bibfnamefont{S.~J.} \bibnamefont{Moss}} \bibnamefont{and}
  \bibinfo{author}{\bibfnamefont{T.}~\bibnamefont{Smart}},
  \bibinfo{journal}{Nat. Rev. Neurosci.} \textbf{\bibinfo{volume}{2}},
  \bibinfo{pages}{240} (\bibinfo{year}{2001}).

\bibitem[{\citenamefont{Sheng and Sala}(2001)}]{ShengSala01}
\bibinfo{author}{\bibfnamefont{M.}~\bibnamefont{Sheng}} \bibnamefont{and}
  \bibinfo{author}{\bibfnamefont{C.}~\bibnamefont{Sala}},
  \bibinfo{journal}{Annu. Rev. Neurosci.} \textbf{\bibinfo{volume}{24}},
  \bibinfo{pages}{1} (\bibinfo{year}{2001}).

\bibitem[{\citenamefont{Faber et~al.}(1985)\citenamefont{Faber, Funch, and
  Korn}}]{Faber85}
\bibinfo{author}{\bibfnamefont{D.~S.} \bibnamefont{Faber}},
  \bibinfo{author}{\bibfnamefont{P.~G.} \bibnamefont{Funch}}, \bibnamefont{and}
  \bibinfo{author}{\bibfnamefont{H.}~\bibnamefont{Korn}},
  \bibinfo{journal}{Proc. Natl. Acad. Sci. U. S. A.}
  \textbf{\bibinfo{volume}{82}}, \bibinfo{pages}{3504} (\bibinfo{year}{1985}).

\bibitem[{\citenamefont{Kullmann and Asztely}(1998)}]{Kullmann98}
\bibinfo{author}{\bibfnamefont{D.~M.} \bibnamefont{Kullmann}} \bibnamefont{and}
  \bibinfo{author}{\bibfnamefont{F.}~\bibnamefont{Asztely}},
  \bibinfo{journal}{Trends Neurosci.} \textbf{\bibinfo{volume}{21}},
  \bibinfo{pages}{8} (\bibinfo{year}{1998}).

\bibitem[{\citenamefont{Clark and {Cull-Candy}}(2002)}]{Clark02}
\bibinfo{author}{\bibfnamefont{B.~A.} \bibnamefont{Clark}} \bibnamefont{and}
  \bibinfo{author}{\bibfnamefont{S.~G.} \bibnamefont{{Cull-Candy}}},
  \bibinfo{journal}{J. Neurosci.} \textbf{\bibinfo{volume}{22}},
  \bibinfo{pages}{4428} (\bibinfo{year}{2002}).

\bibitem[{\citenamefont{Momiyama et~al.}(2003)\citenamefont{Momiyama, Silver,
  H{\"{a}}usser, Notomi, Wu, Shigemoto, and {Cull-Candy}}}]{Momiyama03}
\bibinfo{author}{\bibfnamefont{A.}~\bibnamefont{Momiyama}},
  \bibinfo{author}{\bibfnamefont{R.~A.} \bibnamefont{Silver}},
  \bibinfo{author}{\bibfnamefont{M.}~\bibnamefont{H{\"{a}}usser}},
  \bibinfo{author}{\bibfnamefont{T.}~\bibnamefont{Notomi}},
  \bibinfo{author}{\bibfnamefont{Y.}~\bibnamefont{Wu}},
  \bibinfo{author}{\bibfnamefont{R.}~\bibnamefont{Shigemoto}},
  \bibnamefont{and} \bibinfo{author}{\bibfnamefont{S.~G.}
  \bibnamefont{{Cull-Candy}}}, \bibinfo{journal}{J. Physiol.}
  \textbf{\bibinfo{volume}{549}}, \bibinfo{pages}{75} (\bibinfo{year}{2003}).

\bibitem[{\citenamefont{Scimemi et~al.}(2004)\citenamefont{Scimemi, Fine,
  Kullmann, and Rusakov}}]{kullmann2004}
\bibinfo{author}{\bibfnamefont{A.}~\bibnamefont{Scimemi}},
  \bibinfo{author}{\bibfnamefont{A.}~\bibnamefont{Fine}},
  \bibinfo{author}{\bibfnamefont{D.~M.} \bibnamefont{Kullmann}},
  \bibnamefont{and} \bibinfo{author}{\bibfnamefont{D.~A.}
  \bibnamefont{Rusakov}}, \bibinfo{journal}{J. Neurosc.}
  \textbf{\bibinfo{volume}{24}}, \bibinfo{pages}{4767} (\bibinfo{year}{2004}).

\bibitem[{\citenamefont{Newman}(2003)}]{Newmann03}
\bibinfo{author}{\bibfnamefont{E.~A.} \bibnamefont{Newman}},
  \bibinfo{journal}{Trends Neurosci.} \textbf{\bibinfo{volume}{26}},
  \bibinfo{pages}{536} (\bibinfo{year}{2003}).

\bibitem[{\citenamefont{Chih et~al.}(2005)\citenamefont{Chih, Engelman, and
  Scheiffele}}]{ChihScience05}
\bibinfo{author}{\bibfnamefont{B.}~\bibnamefont{Chih}},
  \bibinfo{author}{\bibfnamefont{H.}~\bibnamefont{Engelman}}, \bibnamefont{and}
  \bibinfo{author}{\bibfnamefont{P.}~\bibnamefont{Scheiffele}},
  \bibinfo{journal}{Science} \textbf{\bibinfo{volume}{307}},
  \bibinfo{pages}{1324} (\bibinfo{year}{2005}).

\bibitem[{\citenamefont{Hussain and Sheng}(2005)}]{Sheng05}
\bibinfo{author}{\bibfnamefont{N.~K.} \bibnamefont{Hussain}} \bibnamefont{and}
  \bibinfo{author}{\bibfnamefont{M.}~\bibnamefont{Sheng}},
  \bibinfo{journal}{Science} \textbf{\bibinfo{volume}{307}},
  \bibinfo{pages}{1207} (\bibinfo{year}{2005}).

\bibitem[{\citenamefont{Choquet and Triller}(2003)}]{DC-AT}
\bibinfo{author}{\bibfnamefont{D.}~\bibnamefont{Choquet}} \bibnamefont{and}
  \bibinfo{author}{\bibfnamefont{A.}~\bibnamefont{Triller}},
  \bibinfo{journal}{Nat. Rev. Neurosci.} \textbf{\bibinfo{volume}{4}},
  \bibinfo{pages}{251} (\bibinfo{year}{2003}).

\bibitem[{\citenamefont{Landau and Lifshitz}(1998)}]{landau-std}
\bibinfo{author}{\bibfnamefont{L.~D.} \bibnamefont{Landau}} \bibnamefont{and}
  \bibinfo{author}{\bibfnamefont{E.~M.} \bibnamefont{Lifshitz}},
  \emph{\bibinfo{title}{Statistical Physics : Part 1 (Course of Theoretical
  Physics, Volume 5)}} (\bibinfo{publisher}{Butterworth-Heinemann},
  \bibinfo{year}{1998}), \bibinfo{edition}{3rd} ed.

\bibitem[{\citenamefont{Allison et~al.}(2000)\citenamefont{Allison, Chervin,
  Gelfand, and Craig}}]{Allison00}
\bibinfo{author}{\bibfnamefont{D.~W.} \bibnamefont{Allison}},
  \bibinfo{author}{\bibfnamefont{A.~S.} \bibnamefont{Chervin}},
  \bibinfo{author}{\bibfnamefont{V.~I.} \bibnamefont{Gelfand}},
  \bibnamefont{and} \bibinfo{author}{\bibfnamefont{A.~M.} \bibnamefont{Craig}},
  \bibinfo{journal}{J. Neurosci.} \textbf{\bibinfo{volume}{20}},
  \bibinfo{pages}{4545} (\bibinfo{year}{2000}).

\bibitem[{\citenamefont{{van Effenterre} and Roux}(2003)}]{Roux05}
\bibinfo{author}{\bibfnamefont{D.}~\bibnamefont{{van Effenterre}}}
  \bibnamefont{and} \bibinfo{author}{\bibfnamefont{D.}~\bibnamefont{Roux}},
  \bibinfo{journal}{Europhys. Lett.} \textbf{\bibinfo{volume}{64}},
  \bibinfo{pages}{543} (\bibinfo{year}{2003}).

\bibitem[{\citenamefont{Smith and Seifert}(2005)}]{Smith-Udo}
\bibinfo{author}{\bibfnamefont{A.~S.} \bibnamefont{Smith}} \bibnamefont{and}
  \bibinfo{author}{\bibfnamefont{U.}~\bibnamefont{Seifert}},
  \bibinfo{journal}{Phys. Rev. E} \textbf{\bibinfo{volume}{71}},
  \bibinfo{pages}{061902} (\bibinfo{year}{2005}).

\bibitem[{\citenamefont{Sarda et~al.}(2004)\citenamefont{Sarda, Pointu, Pincet,
  and Henry}}]{Pincet04}
\bibinfo{author}{\bibfnamefont{S.}~\bibnamefont{Sarda}},
  \bibinfo{author}{\bibfnamefont{D.}~\bibnamefont{Pointu}},
  \bibinfo{author}{\bibfnamefont{F.}~\bibnamefont{Pincet}}, \bibnamefont{and}
  \bibinfo{author}{\bibfnamefont{N.}~\bibnamefont{Henry}},
  \bibinfo{journal}{Biophys. J.} \textbf{\bibinfo{volume}{86}},
  \bibinfo{pages}{3291} (\bibinfo{year}{2004}).

\bibitem[{\citenamefont{Andelman and Joanny}(1993)}]{AndelmanJF}
\bibinfo{author}{\bibfnamefont{D.}~\bibnamefont{Andelman}} \bibnamefont{and}
  \bibinfo{author}{\bibfnamefont{J.~F.} \bibnamefont{Joanny}},
  \bibinfo{journal}{J. Phys. II France} \textbf{\bibinfo{volume}{3}},
  \bibinfo{pages}{121} (\bibinfo{year}{1993}).

\bibitem[{\citenamefont{Derkach et~al.}(2007)\citenamefont{Derkach, Oh, Guire,
  and Soderling}}]{Derkach07}
\bibinfo{author}{\bibfnamefont{V.~A.} \bibnamefont{Derkach}},
  \bibinfo{author}{\bibfnamefont{M.~C.} \bibnamefont{Oh}},
  \bibinfo{author}{\bibfnamefont{E.~S.} \bibnamefont{Guire}}, \bibnamefont{and}
  \bibinfo{author}{\bibfnamefont{T.~R.} \bibnamefont{Soderling}},
  \bibinfo{journal}{Nat. Rev. Neurosci.} \textbf{\bibinfo{volume}{8}},
  \bibinfo{pages}{101} (\bibinfo{year}{2007}).

\bibitem[{\citenamefont{Rasmussen et~al.}(2002)\citenamefont{Rasmussen,
  Rasmussen, Triller, and Vannier}}]{Rasmussen02}
\bibinfo{author}{\bibfnamefont{H.}~\bibnamefont{Rasmussen}},
  \bibinfo{author}{\bibfnamefont{T.}~\bibnamefont{Rasmussen}},
  \bibinfo{author}{\bibfnamefont{A.}~\bibnamefont{Triller}}, \bibnamefont{and}
  \bibinfo{author}{\bibfnamefont{C.}~\bibnamefont{Vannier}},
  \bibinfo{journal}{Mol. Cell Neurosci.} \textbf{\bibinfo{volume}{19}},
  \bibinfo{pages}{201} (\bibinfo{year}{2002}).

\bibitem[{\citenamefont{Okabe et~al.}(2001)\citenamefont{Okabe, Urushido,
  Konno, Okado, and Sobue}}]{temps1}
\bibinfo{author}{\bibfnamefont{S.}~\bibnamefont{Okabe}},
  \bibinfo{author}{\bibfnamefont{T.}~\bibnamefont{Urushido}},
  \bibinfo{author}{\bibfnamefont{D.}~\bibnamefont{Konno}},
  \bibinfo{author}{\bibfnamefont{H.}~\bibnamefont{Okado}}, \bibnamefont{and}
  \bibinfo{author}{\bibfnamefont{K.}~\bibnamefont{Sobue}}, \bibinfo{journal}{J.
  Neurosci} \textbf{\bibinfo{volume}{21}}, \bibinfo{pages}{9561}
  (\bibinfo{year}{2001}).

\bibitem[{\citenamefont{Nakagawa et~al.}(2004)\citenamefont{Nakagawa, Engler,
  and Sheng}}]{temps2}
\bibinfo{author}{\bibfnamefont{T.}~\bibnamefont{Nakagawa}},
  \bibinfo{author}{\bibfnamefont{J.~A.} \bibnamefont{Engler}},
  \bibnamefont{and} \bibinfo{author}{\bibfnamefont{M.}~\bibnamefont{Sheng}},
  \bibinfo{journal}{Neuropharmacology} \textbf{\bibinfo{volume}{47}},
  \bibinfo{pages}{734} (\bibinfo{year}{2004}).

\bibitem[{\citenamefont{Star et~al.}(2002)\citenamefont{Star, Kwiatkowski, and
  Murthy}}]{Star02}
\bibinfo{author}{\bibfnamefont{E.~N.} \bibnamefont{Star}},
  \bibinfo{author}{\bibfnamefont{D.~J.} \bibnamefont{Kwiatkowski}},
  \bibnamefont{and} \bibinfo{author}{\bibfnamefont{V.~N.}
  \bibnamefont{Murthy}}, \bibinfo{journal}{Nat. Neurosci.}
  \textbf{\bibinfo{volume}{5}}, \bibinfo{pages}{239} (\bibinfo{year}{2002}).

\bibitem[{\citenamefont{Sieber et~al.}(2007)\citenamefont{Sieber, Willig,
  Kutzner, {Gerding-Reimers}, Harke, Donnert, Rammner, Eggeling, Hell,
  {Grubm\"{u}ller} et~al.}}]{Size-Lang-Sci07}
\bibinfo{author}{\bibfnamefont{J.~J.} \bibnamefont{Sieber}},
  \bibinfo{author}{\bibfnamefont{K.~I.} \bibnamefont{Willig}},
  \bibinfo{author}{\bibfnamefont{C.}~\bibnamefont{Kutzner}},
  \bibinfo{author}{\bibfnamefont{C.}~\bibnamefont{{Gerding-Reimers}}},
  \bibinfo{author}{\bibfnamefont{B.}~\bibnamefont{Harke}},
  \bibinfo{author}{\bibfnamefont{G.}~\bibnamefont{Donnert}},
  \bibinfo{author}{\bibfnamefont{B.}~\bibnamefont{Rammner}},
  \bibinfo{author}{\bibfnamefont{C.}~\bibnamefont{Eggeling}},
  \bibinfo{author}{\bibfnamefont{S.~W.} \bibnamefont{Hell}},
  \bibinfo{author}{\bibfnamefont{H.}~\bibnamefont{{Grubm\"{u}ller}}},
  \bibnamefont{et~al.}, \bibinfo{journal}{Science}
  \textbf{\bibinfo{volume}{317}}, \bibinfo{pages}{1072} (\bibinfo{year}{2007}).

\bibitem[{\citenamefont{Bedet et~al.}(2006)\citenamefont{Bedet, Bruusgaard,
  Vergo, {Groth-Pedersen}, Eimer, Triller, and Vannier}}]{Bedet-JBC06}
\bibinfo{author}{\bibfnamefont{C.}~\bibnamefont{Bedet}},
  \bibinfo{author}{\bibfnamefont{J.~C.} \bibnamefont{Bruusgaard}},
  \bibinfo{author}{\bibfnamefont{S.}~\bibnamefont{Vergo}},
  \bibinfo{author}{\bibfnamefont{L.}~\bibnamefont{{Groth-Pedersen}}},
  \bibinfo{author}{\bibfnamefont{S.}~\bibnamefont{Eimer}},
  \bibinfo{author}{\bibfnamefont{A.}~\bibnamefont{Triller}}, \bibnamefont{and}
  \bibinfo{author}{\bibfnamefont{C.}~\bibnamefont{Vannier}},
  \bibinfo{journal}{J. Biol. Chem.} \textbf{\bibinfo{volume}{281}},
  \bibinfo{pages}{30046} (\bibinfo{year}{2006}).

\bibitem[{\citenamefont{Fritschy et~al.}(2008)\citenamefont{Fritschy, Harvey,
  and Schwarze}}]{gephyerin-TINS08}
\bibinfo{author}{\bibfnamefont{J.~M.} \bibnamefont{Fritschy}},
  \bibinfo{author}{\bibfnamefont{R.~J.} \bibnamefont{Harvey}},
  \bibnamefont{and} \bibinfo{author}{\bibfnamefont{G.}~\bibnamefont{Schwarze}},
  \bibinfo{journal}{Trends Neurosci.} \textbf{\bibinfo{volume}{31}},
  \bibinfo{pages}{257} (\bibinfo{year}{2008}).

\bibitem[{\citenamefont{Sheng and Hoogenraad}(2006)}]{Sheng-AnnuRev07}
\bibinfo{author}{\bibfnamefont{M.}~\bibnamefont{Sheng}} \bibnamefont{and}
  \bibinfo{author}{\bibfnamefont{C.~C.} \bibnamefont{Hoogenraad}},
  \bibinfo{journal}{Anuu. Rev. Biochem.} \textbf{\bibinfo{volume}{76}},
  \bibinfo{pages}{1} (\bibinfo{year}{2006}).

\bibitem[{\citenamefont{{Masugi-Tokita}
  et~al.}(2007)\citenamefont{{Masugi-Tokita}, Tarusawa, Watanabe, {Moln\'ar},
  Fujimoto, and Shigemoto}}]{shigemoto07}
\bibinfo{author}{\bibfnamefont{M.}~\bibnamefont{{Masugi-Tokita}}},
  \bibinfo{author}{\bibfnamefont{E.}~\bibnamefont{Tarusawa}},
  \bibinfo{author}{\bibfnamefont{M.}~\bibnamefont{Watanabe}},
  \bibinfo{author}{\bibfnamefont{E.}~\bibnamefont{{Moln\'ar}}},
  \bibinfo{author}{\bibfnamefont{K.}~\bibnamefont{Fujimoto}}, \bibnamefont{and}
  \bibinfo{author}{\bibfnamefont{R.}~\bibnamefont{Shigemoto}},
  \bibinfo{journal}{J. Neurosci.} \textbf{\bibinfo{volume}{27}},
  \bibinfo{pages}{2135} (\bibinfo{year}{2007}).

\bibitem[{\citenamefont{Chen et~al.}(2005)\citenamefont{Chen, Vinade, Leapman,
  Petersen, Nakagawa, Phillips, Sheng, and Reese}}]{PSdensity-Sheng}
\bibinfo{author}{\bibfnamefont{X.}~\bibnamefont{Chen}},
  \bibinfo{author}{\bibfnamefont{L.}~\bibnamefont{Vinade}},
  \bibinfo{author}{\bibfnamefont{R.~D.} \bibnamefont{Leapman}},
  \bibinfo{author}{\bibfnamefont{J.~D.} \bibnamefont{Petersen}},
  \bibinfo{author}{\bibfnamefont{T.}~\bibnamefont{Nakagawa}},
  \bibinfo{author}{\bibfnamefont{T.~M.} \bibnamefont{Phillips}},
  \bibinfo{author}{\bibfnamefont{M.}~\bibnamefont{Sheng}}, \bibnamefont{and}
  \bibinfo{author}{\bibfnamefont{T.~S.} \bibnamefont{Reese}},
  \bibinfo{journal}{Proc. Natl. Acad. Sci. U. S. A.}
  \textbf{\bibinfo{volume}{102}}, \bibinfo{pages}{11551}
  (\bibinfo{year}{2005}).

\bibitem[{\citenamefont{Kennedy}(2000)}]{Signal-Kennedy}
\bibinfo{author}{\bibfnamefont{M.~B.} \bibnamefont{Kennedy}},
  \bibinfo{journal}{Science} \textbf{\bibinfo{volume}{290}},
  \bibinfo{pages}{750} (\bibinfo{year}{2000}).

\bibitem[{\citenamefont{Zita et~al.}(2007)\citenamefont{Zita, Marchionni,
  Bottos, Righi, Sal, Cherubini, and Zacchi}}]{phosphoryl-Zeta-EMBO07}
\bibinfo{author}{\bibfnamefont{M.~M.} \bibnamefont{Zita}},
  \bibinfo{author}{\bibfnamefont{I.}~\bibnamefont{Marchionni}},
  \bibinfo{author}{\bibfnamefont{E.}~\bibnamefont{Bottos}},
  \bibinfo{author}{\bibfnamefont{M.}~\bibnamefont{Righi}},
  \bibinfo{author}{\bibfnamefont{G.~D.} \bibnamefont{Sal}},
  \bibinfo{author}{\bibfnamefont{E.}~\bibnamefont{Cherubini}},
  \bibnamefont{and} \bibinfo{author}{\bibfnamefont{P.}~\bibnamefont{Zacchi}},
  \bibinfo{journal}{EMBO J.} \textbf{\bibinfo{volume}{26}},
  \bibinfo{pages}{1761} (\bibinfo{year}{2007}).

\bibitem[{\citenamefont{Misteli}(2001)}]{Misteli}
\bibinfo{author}{\bibfnamefont{T.}~\bibnamefont{Misteli}}, \bibinfo{journal}{J.
  Cell. Biol.} \textbf{\bibinfo{volume}{155}}, \bibinfo{pages}{181}
  (\bibinfo{year}{2001}).

\bibitem[{\citenamefont{Shouval}(2005)}]{shouval}
\bibinfo{author}{\bibfnamefont{H.~Z.} \bibnamefont{Shouval}},
  \bibinfo{journal}{Proc. Natl. Acad. Sci. U. S. A.}
  \textbf{\bibinfo{volume}{102}}, \bibinfo{pages}{14440}
  (\bibinfo{year}{2005}).

\bibitem[{\citenamefont{Yeramian and {J.-P. Changeux}}(1986)}]{Changeux}
\bibinfo{author}{\bibfnamefont{E.}~\bibnamefont{Yeramian}} \bibnamefont{and}
  \bibinfo{author}{\bibnamefont{{J.-P. Changeux}}}, \bibinfo{journal}{C. R.
  Acad. Sci. III} \textbf{\bibinfo{volume}{302}}, \bibinfo{pages}{609}
  (\bibinfo{year}{1986}).

\bibitem[{\citenamefont{Fusi et~al.}(2005)\citenamefont{Fusi, Drew, and
  Abbott}}]{switch}
\bibinfo{author}{\bibfnamefont{S.}~\bibnamefont{Fusi}},
  \bibinfo{author}{\bibfnamefont{P.~J.} \bibnamefont{Drew}}, \bibnamefont{and}
  \bibinfo{author}{\bibfnamefont{L.}~\bibnamefont{Abbott}},
  \bibinfo{journal}{Neuron} \textbf{\bibinfo{volume}{45}}, \bibinfo{pages}{599}
  (\bibinfo{year}{2005}).

\bibitem[{\citenamefont{Smolen}(1981)}]{h0devel-BrRes81}
\bibinfo{author}{\bibfnamefont{A.~J.} \bibnamefont{Smolen}},
  \bibinfo{journal}{Brain Res.} \textbf{\bibinfo{volume}{227}},
  \bibinfo{pages}{49} (\bibinfo{year}{1981}).

\bibitem[{\citenamefont{Gentschev and Sotelo}(1973)}]{sotelo-brain-res73}
\bibinfo{author}{\bibfnamefont{T.}~\bibnamefont{Gentschev}} \bibnamefont{and}
  \bibinfo{author}{\bibfnamefont{C.}~\bibnamefont{Sotelo}},
  \bibinfo{journal}{Brain Res.} \textbf{\bibinfo{volume}{62}},
  \bibinfo{pages}{37} (\bibinfo{year}{1973}).

\bibitem[{\citenamefont{Seitanidou et~al.}(1992)\citenamefont{Seitanidou,
  Nicola, Triller, and Korn}}]{denervation-AT-JN92}
\bibinfo{author}{\bibfnamefont{T.}~\bibnamefont{Seitanidou}},
  \bibinfo{author}{\bibfnamefont{M.~A.} \bibnamefont{Nicola}},
  \bibinfo{author}{\bibfnamefont{A.}~\bibnamefont{Triller}}, \bibnamefont{and}
  \bibinfo{author}{\bibfnamefont{H.}~\bibnamefont{Korn}}, \bibinfo{journal}{J.
  Neurosci.} \textbf{\bibinfo{volume}{12}}, \bibinfo{pages}{116}
  (\bibinfo{year}{1992}).

\bibitem[{\citenamefont{Friedman et~al.}(2000)\citenamefont{Friedman, Bresler,
  Garner, and Ziv}}]{Friedman-neuron00}
\bibinfo{author}{\bibfnamefont{H.~V.} \bibnamefont{Friedman}},
  \bibinfo{author}{\bibfnamefont{T.}~\bibnamefont{Bresler}},
  \bibinfo{author}{\bibfnamefont{C.~C.} \bibnamefont{Garner}},
  \bibnamefont{and} \bibinfo{author}{\bibfnamefont{N.~E.} \bibnamefont{Ziv}},
  \bibinfo{journal}{Neuron} \textbf{\bibinfo{volume}{27}}, \bibinfo{pages}{57}
  (\bibinfo{year}{2000}).

\bibitem[{\citenamefont{Scheiffele}(2003)}]{Scheiffele03}
\bibinfo{author}{\bibfnamefont{P.}~\bibnamefont{Scheiffele}},
  \bibinfo{journal}{Annu. Rev. Neurosci.} \textbf{\bibinfo{volume}{26}},
  \bibinfo{pages}{485} (\bibinfo{year}{2003}).

\bibitem[{\citenamefont{Basarsky et~al.}(1994)\citenamefont{Basarsky, Parpura,
  and Haydon}}]{Basarsky94}
\bibinfo{author}{\bibfnamefont{T.~A.} \bibnamefont{Basarsky}},
  \bibinfo{author}{\bibfnamefont{V.}~\bibnamefont{Parpura}}, \bibnamefont{and}
  \bibinfo{author}{\bibfnamefont{P.~G.} \bibnamefont{Haydon}},
  \bibinfo{journal}{J. Neurosci.} \textbf{\bibinfo{volume}{14}},
  \bibinfo{pages}{6402} (\bibinfo{year}{1994}).

\bibitem[{\citenamefont{Colin et~al.}(1996)\citenamefont{Colin, Rostaing, and
  Triller}}]{Colin-JCompara96}
\bibinfo{author}{\bibfnamefont{I.}~\bibnamefont{Colin}},
  \bibinfo{author}{\bibfnamefont{P.}~\bibnamefont{Rostaing}}, \bibnamefont{and}
  \bibinfo{author}{\bibfnamefont{A.}~\bibnamefont{Triller}},
  \bibinfo{journal}{J. Comp. Neurol.} \textbf{\bibinfo{volume}{374}},
  \bibinfo{pages}{467} (\bibinfo{year}{1996}).

\bibitem[{\citenamefont{Colin et~al.}(1998)\citenamefont{Colin, Rostaing,
  Augustin, and Triller}}]{Colin-JCompara98}
\bibinfo{author}{\bibfnamefont{I.}~\bibnamefont{Colin}},
  \bibinfo{author}{\bibfnamefont{P.}~\bibnamefont{Rostaing}},
  \bibinfo{author}{\bibfnamefont{A.}~\bibnamefont{Augustin}}, \bibnamefont{and}
  \bibinfo{author}{\bibfnamefont{A.}~\bibnamefont{Triller}},
  \bibinfo{journal}{J. Comp. Neurol.} \textbf{\bibinfo{volume}{398}},
  \bibinfo{pages}{359} (\bibinfo{year}{1998}).

\bibitem[{\citenamefont{Dityatev and
  Schachner}(2003)}]{extracell-Dityatev-Nature03}
\bibinfo{author}{\bibfnamefont{A.}~\bibnamefont{Dityatev}} \bibnamefont{and}
  \bibinfo{author}{\bibfnamefont{M.}~\bibnamefont{Schachner}},
  \bibinfo{journal}{Nat. Rev. Neurosci.} \textbf{\bibinfo{volume}{4}},
  \bibinfo{pages}{456} (\bibinfo{year}{2003}).

\bibitem[{\citenamefont{Sylantyev et~al.}(2008)\citenamefont{Sylantyev,
  Savtchenko, Niu, Ivanov, Jensen, Kullmann, Xiao, and
  Rusakov}}]{Efields-Sylantyev-Science08}
\bibinfo{author}{\bibfnamefont{S.}~\bibnamefont{Sylantyev}},
  \bibinfo{author}{\bibfnamefont{L.~P.} \bibnamefont{Savtchenko}},
  \bibinfo{author}{\bibfnamefont{Y.}~\bibnamefont{Niu}},
  \bibinfo{author}{\bibfnamefont{A.~I.} \bibnamefont{Ivanov}},
  \bibinfo{author}{\bibfnamefont{T.~P.} \bibnamefont{Jensen}},
  \bibinfo{author}{\bibfnamefont{D.~M.} \bibnamefont{Kullmann}},
  \bibinfo{author}{\bibfnamefont{M.-Y.} \bibnamefont{Xiao}}, \bibnamefont{and}
  \bibinfo{author}{\bibfnamefont{D.~A.} \bibnamefont{Rusakov}},
  \bibinfo{journal}{Science} \textbf{\bibinfo{volume}{319}},
  \bibinfo{pages}{1845} (\bibinfo{year}{2008}).

\bibitem[{\citenamefont{Matsuzaki et~al.}(2001)\citenamefont{Matsuzaki,
  {Ellis-Davies}, Nemoto, Miyashita, Iino, and
  Kasai}}]{spine-geometry-NatNeur01}
\bibinfo{author}{\bibfnamefont{M.}~\bibnamefont{Matsuzaki}},
  \bibinfo{author}{\bibfnamefont{G.~C.} \bibnamefont{{Ellis-Davies}}},
  \bibinfo{author}{\bibfnamefont{T.}~\bibnamefont{Nemoto}},
  \bibinfo{author}{\bibfnamefont{Y.}~\bibnamefont{Miyashita}},
  \bibinfo{author}{\bibfnamefont{M.}~\bibnamefont{Iino}}, \bibnamefont{and}
  \bibinfo{author}{\bibfnamefont{H.}~\bibnamefont{Kasai}},
  \bibinfo{journal}{Nat. Neurosci.} \textbf{\bibinfo{volume}{4}},
  \bibinfo{pages}{1086} (\bibinfo{year}{2001}).

\bibitem[{\citenamefont{Rasse et~al.}(2005)\citenamefont{Rasse, Fouquet,
  Schmid, Kittel, Mertel, Sigrist, Schmidt, Guzman, Merino, Qin
  et~al.}}]{Glu-Dyn-Rasse}
\bibinfo{author}{\bibfnamefont{T.~M.} \bibnamefont{Rasse}},
  \bibinfo{author}{\bibfnamefont{W.}~\bibnamefont{Fouquet}},
  \bibinfo{author}{\bibfnamefont{A.}~\bibnamefont{Schmid}},
  \bibinfo{author}{\bibfnamefont{R.~J.} \bibnamefont{Kittel}},
  \bibinfo{author}{\bibfnamefont{S.}~\bibnamefont{Mertel}},
  \bibinfo{author}{\bibfnamefont{C.~B.} \bibnamefont{Sigrist}},
  \bibinfo{author}{\bibfnamefont{M.}~\bibnamefont{Schmidt}},
  \bibinfo{author}{\bibfnamefont{A.}~\bibnamefont{Guzman}},
  \bibinfo{author}{\bibfnamefont{C.}~\bibnamefont{Merino}},
  \bibinfo{author}{\bibfnamefont{G.}~\bibnamefont{Qin}}, \bibnamefont{et~al.},
  \bibinfo{journal}{Nat. Neurosci.} \textbf{\bibinfo{volume}{8}},
  \bibinfo{pages}{898} (\bibinfo{year}{2005}).

\bibitem[{\citenamefont{Peters et~al.}(1991)\citenamefont{Peters, Palay, and
  {H. deF. Webster}}}]{Peters76}
\bibinfo{author}{\bibfnamefont{A.}~\bibnamefont{Peters}},
  \bibinfo{author}{\bibfnamefont{S.~L.} \bibnamefont{Palay}}, \bibnamefont{and}
  \bibinfo{author}{\bibnamefont{{H. deF. Webster}}}, \emph{\bibinfo{title}{Fine
  Structure of the Nervous System. {N}eurons and their Supporting Cells}}
  (\bibinfo{publisher}{Oxford University Press, U.S.A.}, \bibinfo{year}{1991}),
  \bibinfo{edition}{3rd} ed.

\bibitem[{\citenamefont{Meier et~al.}(2000)\citenamefont{Meier,
  {Meunier-Durmort}, Forest, Triller, and Vannier}}]{Meier-JCellS00}
\bibinfo{author}{\bibfnamefont{J.}~\bibnamefont{Meier}},
  \bibinfo{author}{\bibfnamefont{C.}~\bibnamefont{{Meunier-Durmort}}},
  \bibinfo{author}{\bibfnamefont{C.}~\bibnamefont{Forest}},
  \bibinfo{author}{\bibfnamefont{A.}~\bibnamefont{Triller}}, \bibnamefont{and}
  \bibinfo{author}{\bibfnamefont{C.}~\bibnamefont{Vannier}},
  \bibinfo{journal}{J. Cell. Sci.} \textbf{\bibinfo{volume}{113}},
  \bibinfo{pages}{2783} (\bibinfo{year}{2000}).

\bibitem[{\citenamefont{Gilbert et~al.}(1997)\citenamefont{Gilbert, Paccaud,
  and Carpentier}}]{clathrin_JCS97}
\bibinfo{author}{\bibfnamefont{A.}~\bibnamefont{Gilbert}},
  \bibinfo{author}{\bibfnamefont{J.~P.} \bibnamefont{Paccaud}},
  \bibnamefont{and} \bibinfo{author}{\bibfnamefont{J.~L.}
  \bibnamefont{Carpentier}}, \bibinfo{journal}{J. Cell. Sci.}
  \textbf{\bibinfo{volume}{110}}, \bibinfo{pages}{3105} (\bibinfo{year}{1997}).

\bibitem[{\citenamefont{Sheng and Nakagawa}(2002)}]{ShengParadox}
\bibinfo{author}{\bibfnamefont{M.}~\bibnamefont{Sheng}} \bibnamefont{and}
  \bibinfo{author}{\bibfnamefont{T.}~\bibnamefont{Nakagawa}},
  \bibinfo{journal}{Nature} \textbf{\bibinfo{volume}{417}},
  \bibinfo{pages}{601} (\bibinfo{year}{2002}).

\bibitem[{\citenamefont{Bresler et~al.}(2004)\citenamefont{Bresler, Shapira,
  Boeckers, Dresbach, Futter, Garner, Rosenblum, Gundelfinger, and
  Ziv}}]{Bresler04}
\bibinfo{author}{\bibfnamefont{T.}~\bibnamefont{Bresler}},
  \bibinfo{author}{\bibfnamefont{M.}~\bibnamefont{Shapira}},
  \bibinfo{author}{\bibfnamefont{T.}~\bibnamefont{Boeckers}},
  \bibinfo{author}{\bibfnamefont{T.}~\bibnamefont{Dresbach}},
  \bibinfo{author}{\bibfnamefont{M.}~\bibnamefont{Futter}},
  \bibinfo{author}{\bibfnamefont{C.~C.} \bibnamefont{Garner}},
  \bibinfo{author}{\bibfnamefont{K.}~\bibnamefont{Rosenblum}},
  \bibinfo{author}{\bibfnamefont{E.~D.} \bibnamefont{Gundelfinger}},
  \bibnamefont{and} \bibinfo{author}{\bibfnamefont{N.~E.} \bibnamefont{Ziv}},
  \bibinfo{journal}{J. Neurosci} \textbf{\bibinfo{volume}{24}},
  \bibinfo{pages}{1507} (\bibinfo{year}{2004}).

\bibitem[{\citenamefont{Ehrensperger et~al.}(2007)\citenamefont{Ehrensperger,
  Hanus, Vannier, Triller, and Dahan}}]{Erensperger-BJ07}
\bibinfo{author}{\bibfnamefont{M.~V.} \bibnamefont{Ehrensperger}},
  \bibinfo{author}{\bibfnamefont{C.}~\bibnamefont{Hanus}},
  \bibinfo{author}{\bibfnamefont{C.}~\bibnamefont{Vannier}},
  \bibinfo{author}{\bibfnamefont{A.}~\bibnamefont{Triller}}, \bibnamefont{and}
  \bibinfo{author}{\bibfnamefont{M.}~\bibnamefont{Dahan}},
  \bibinfo{journal}{Biophys. J.} \textbf{\bibinfo{volume}{92}},
  \bibinfo{pages}{3706} (\bibinfo{year}{2007}).

\bibitem[{\citenamefont{{A. E. El-Husseini} et~al.}(2000)\citenamefont{{A. E.
  El-Husseini}, Craven, Chetkovich, Firestein, Schnell, Aoki, and
  Bredt}}]{El-Husseini1-JCB00}
\bibinfo{author}{\bibnamefont{{A. E. El-Husseini}}},
  \bibinfo{author}{\bibfnamefont{S.~E.} \bibnamefont{Craven}},
  \bibinfo{author}{\bibfnamefont{D.~M.} \bibnamefont{Chetkovich}},
  \bibinfo{author}{\bibfnamefont{B.~L.} \bibnamefont{Firestein}},
  \bibinfo{author}{\bibfnamefont{E.}~\bibnamefont{Schnell}},
  \bibinfo{author}{\bibfnamefont{C.}~\bibnamefont{Aoki}}, \bibnamefont{and}
  \bibinfo{author}{\bibfnamefont{D.~S.} \bibnamefont{Bredt}},
  \bibinfo{journal}{J. Cell. Biol.} \textbf{\bibinfo{volume}{148}},
  \bibinfo{pages}{159} (\bibinfo{year}{2000}).

\bibitem[{\citenamefont{{Ael-D. El-Husseini} et~al.}(2002)\citenamefont{{Ael-D.
  El-Husseini}, Schnell, Dakoji, Sweeney, Zhou, Prange, {Gauthier-Campbell},
  {Aguilera-Moreno}, Nicoll, and Bredt}}]{El-Husseini1-Cell02}
\bibinfo{author}{\bibnamefont{{Ael-D. El-Husseini}}},
  \bibinfo{author}{\bibfnamefont{E.}~\bibnamefont{Schnell}},
  \bibinfo{author}{\bibfnamefont{S.}~\bibnamefont{Dakoji}},
  \bibinfo{author}{\bibfnamefont{N.}~\bibnamefont{Sweeney}},
  \bibinfo{author}{\bibfnamefont{Q.}~\bibnamefont{Zhou}},
  \bibinfo{author}{\bibfnamefont{O.}~\bibnamefont{Prange}},
  \bibinfo{author}{\bibfnamefont{C.}~\bibnamefont{{Gauthier-Campbell}}},
  \bibinfo{author}{\bibfnamefont{A.}~\bibnamefont{{Aguilera-Moreno}}},
  \bibinfo{author}{\bibfnamefont{R.~A.} \bibnamefont{Nicoll}},
  \bibnamefont{and} \bibinfo{author}{\bibfnamefont{D.~S.} \bibnamefont{Bredt}},
  \bibinfo{journal}{Cell} \textbf{\bibinfo{volume}{108}}, \bibinfo{pages}{849}
  (\bibinfo{year}{2002}).

\bibitem[{\citenamefont{Hanus et~al.}(2004{\natexlab{a}})\citenamefont{Hanus,
  Vannier, and Triller}}]{Cyril-JNS04}
\bibinfo{author}{\bibfnamefont{C.}~\bibnamefont{Hanus}},
  \bibinfo{author}{\bibfnamefont{C.}~\bibnamefont{Vannier}}, \bibnamefont{and}
  \bibinfo{author}{\bibfnamefont{A.}~\bibnamefont{Triller}},
  \bibinfo{journal}{J. Neurosci.} \textbf{\bibinfo{volume}{24}},
  \bibinfo{pages}{1119} (\bibinfo{year}{2004}{\natexlab{a}}).

\bibitem[{\citenamefont{Hanus et~al.}(2004{\natexlab{b}})\citenamefont{Hanus,
  Vannier, and Triller}}]{cyrilAT-JN04}
\bibinfo{author}{\bibfnamefont{C.}~\bibnamefont{Hanus}},
  \bibinfo{author}{\bibfnamefont{C.}~\bibnamefont{Vannier}}, \bibnamefont{and}
  \bibinfo{author}{\bibfnamefont{A.}~\bibnamefont{Triller}},
  \bibinfo{journal}{J. Neurosci.} \textbf{\bibinfo{volume}{24}},
  \bibinfo{pages}{1119} (\bibinfo{year}{2004}{\natexlab{b}}).

\bibitem[{\citenamefont{Triller and Choquet}(2008)}]{ATCDrev08}
\bibinfo{author}{\bibfnamefont{A.}~\bibnamefont{Triller}} \bibnamefont{and}
  \bibinfo{author}{\bibfnamefont{D.}~\bibnamefont{Choquet}},
  \bibinfo{journal}{Neuron} \textbf{\bibinfo{volume}{59}}, \bibinfo{pages}{359}
  (\bibinfo{year}{2008}).

\end{thebibliography}

\end{document}